# TRANSITION PROBABILITIES OF Co II, WEAK LINES TO THE GROUND AND LOW METASTABLE LEVELS

J. E. Lawler[1], T. Feigenson[1], C. Sneden[2], J. J. Cowan[3], and G. Nave[4]

[1]Department of Physics, University of Wisconsin-Madison, 1150 University Ave, Madison, WI 53706; jelawler@wisc.edu; tom.feigenson@gmail.com

[2]Department of Astronomy and McDonald Observatory, University of Texas, Austin, TX 78712; chris@verdi.as.utexas.edu

[3]Homer L. Dodge Department of Physics and Astronomy, University of Oklahoma, Norman, OK 73019; jjcowan1@ou.edu

[4]National Institute of Standards and Technology, 100 Bureau Dr, Gaithersburg, MD 20899; gnave@nist.gov

Abstract

New branching fraction measurements based primarily on data from a cross dispersed echelle spectrometer are reported for 84 lines of Co II. The branching fractions for 82 lines are converted to absolute atomic transition probabilities using radiative lifetimes from laser induced fluorescence (LIF) measurements on 19 upper levels of the lines. A lifetime of 3.3(2) ns for the $z^5D_0$ level is used based on LIF measurements for lifetimes of the four other levels in the $z^5D$ term. Twelve of the 84 lines are weak transitions connecting to the ground and low metastable levels of $Co^+$. Another 46 lines are strong transitions connecting to the ground and low metastable levels of $Co^+$. For these lines log(*gf*) values were measured in earlier studies and, with a few exceptions, are confirmed in this study. Such lines, if unblended in stellar spectra, have the potential to yield Co abundance values unaffected by any breakdown of the Local Thermodynamic Equilibrium approximation in stellar photospheres because the ground and low metastable levels of $Co^+$ are the primary population reservoirs of Co in photospheres of interest. Weak lines, if unblended, are useful in photospheres with high Co abundance and strong lines are useful in metal-poor photospheres. New hyperfine structure A constants for 28 levels of ionized Co from least square fits to Fourier transform spectra line profiles are reported. These laboratory data are applied to re-determine the Co abundance in metal-poor halo star HD 84937. Branching fractions and transition probabilities for 19 lines are reported for the first time.

1. INTRODUCTION

Efforts are underway to map the relative Fe-group elemental abundances in metal-poor (MP) stars. These efforts are motivated by the goals of more quantitative data on the Galactic chemical evolution and a better understanding of the role of early nucleo-synthetic events now being observed as Gamma Ray Bursts (GRBs). Long GRBs are now thought to be early Core Collapse (CC) Super Novae (SNe) from the first generation of stars. Models indicate that the first, zero metallicity stars were massive and likely rapidly rotating (e.g. Abel et al. 2002, Szécsi et al. 2015). These early CC SNe contributed to the Fe-group abundance in MP stars. There is evidence that the relative Fe-group synthesis was different at low metallicity (Henry et al. 2010, Sneden et al. 2016). The differences cannot be explained simply by the rise of the Type I SNe. Short, < 2 s, GRBs are thought to be distant (intermediate redshift) *n*(*eutron*)-star mergers. The *n*-star mergers may contribute some Fe-group material, but they are of greatest interest as a source of *r*(*apid*)-process *n*-capture isotopes.

Work on the relative Fe-group abundances in MP stars is hindered to some extent by limits of Local Thermal Equilibrium (LTE) and One Dimensional (1D) approximations of traditional photospheric models. Recently improved lab data for Cr I (Sobeck et al. 2007), Cr II (Lawler et al. 2017), Mn I & Mn II (Den Hartog et al. 2011), Ti I (Lawler et al. 2013), Ti II (Wood et al. 2013), Ni I (Wood et al. 2014a), V I (Lawler et al. 2014), V II (Wood et al. 2014b), Co I (Lawler et al. 2015), and Fe I (Den Hartog et al. 2014, Ruffoni et al. 2014 & Belmonte et al. 2017) along with applications to the MP dwarf star HD 84937 revealed only a few problems with standard 1D-LTE models (Sneden et al. 2016), but MP giant stars may prove to be more difficult. The electron pressure in MP atmospheres is suppressed in both dwarf and giant stars because most the free electrons are from metals in the F, G, and K stars of interest. This reduced electron pressure can hinder the approach to LTE. Giant stars are essential to elemental abundance surveys on MP stars because they yield the high intrinsic brightness required for high spectral resolution, high signal-to-noise (S/N) spectroscopy. MP giant stars have low density atmospheres and have even lower electron pressures than MP dwarf stars as well as a much lower H pressure. These issues are explored in a review by Asplund (2005). The microturbulent velocity used in standard 1D-LTE photospheric models captures some of the effects of convection, but it is not a substitute for a true 3D model. The most serious challenge of Non-LTE (NLTE) models is the lack of reliable cross sections or rates for inelastic collisions of H atoms with metal atoms and ions (Asplund 2005). Ultimately it will be necessary to compute much better cross sections and rate constants for inelastic and superelastic heavy particle collisions to replace the widely used, but inaccurate, Drawin (1968, 1969) approximation. Barklem (2016) and Barklem et al. (2005, 2011) have made important progress on such cross sections and rate constants. Charge exchange reactions such as $H^- + Fe^+ \leftrightarrow H +$ Fe* are expected to equilibrate highly excited levels of neutral Fe* with the ground and low metastable levels of $Fe^+$. The cryogenic electrostatic storage ring DESIREE at Stockholm University in Sweden will provide accurate, absolute cross sections for important charge exchange reactions at realistic thermal interaction energies using merged beams (Schmidt et al. 2013). Reliable rate constants for other important reactions are also needed. The first 3D/NLTE models are now operational (Amarsi er al. 2017) and the rapidly improving collisional data should make such models highly realistic in the future.

Earlier papers from our collaboration have introduced another approach to this issue of NLTE and 3D effects in metal-poor stars. Lines connected to the ground level and low metastable levels (if such metastable levels exist) of the Fe-group ions, especially unsaturated lines, were described as "Gold Standard Lines". To avoid confusion with an *n*-capture element, we shall henceforth refer to such lines as Highly Reliable Lines (HRLs). These HRLs sample the primary population reservoir levels of Fe-group elements in photospheres of interest. Although low populations of higher levels of a Fe-group ion and any level of a Fe-group neutral atom might be susceptible to NLTE effects, the ground and low metastable levels have most (> 75 %) of the Fe-group material in the photosphere. Singly ionized Co is typical of Fe-group ions. The Saha balance in a photosphere favors the ion over the neutral by ≈ 2 dex (dex = decimal exponent, so 2 dex is a factor of $10^2$) eg, Figure 2 of Sneden et al. (2016). The Atomic Spectra Database (ASD) at the National Institute of Standards and Technology (NIST) has a convenient Partition Function evaluator to combine with Boltzmann factors in a specific ionization stage[5]. Calculations with this tool suggest that $Co^+$ in a photosphere is concentrated in levels of the two lowest terms (< 2/3 eV above the ground level of the ion). At $k_BT = 0.5$ eV and in LTE levels of the two lowest terms have > 90 % of the Co.

Levels that serve as the primary population reservoir for a Fe-group element in a photosphere cannot be pulled significantly out of LTE and excess densities of doubly or multiply ionized Fe-group material cannot persist in a photosphere composed primarily of neutral H due to charge exchange. Multiply ionized Fe-group species have strongly exothermic charge exchange reaction channels with neutral H atoms. Rates constants for such reactions tend to be large due to the lack of potential curve barriers. If a photosphere is composed mostly of H atoms and the rate constants are large, then densities of multiply ionized Fe-group species are maintained at low LTE values.

---

[5] http://physics.nist.gov/PhysRefData/ASD/lines_form.html

For Fe-group abundance measurements on sufficiently MP stars it is possible to determine reliable transition data for UV lines that are dominant branches from upper levels with laser-induced-fluorescence (LIF) lifetime measurements. Transition probabilities of dominant branches tend to have the small, e.g. few percent, uncertainty of a LIF lifetime measurement. For Fe-group abundance measurements on stars with higher (e.g. near Solar) metallicity, it is necessary to use weaker lines to avoid saturation of the absorption features. Although saturation effects can be avoided by choosing lines with high excitation potential (EP), then the issue of possible NLTE effects becomes more of a concern. It is better to choose a line with a small transition probability that connects to the ground and/or low metastable levels. Unfortunately the transition probabilities of minor branches are difficult to measure. By definition branching fractions (BFs) sum to unity, and thus uncertainty migrates to weak branches. The uncertainty can be from the relative radiometric calibration of the BF experiment, from optical depth, from line blends, or other effects such as S/N limits. Transition probabilities for minor branches are also difficult to calculate using ab-initio methods because they are sensitive to level mixing.

Fourier transform spectrometers (FTSs) have been extremely productive in BF measurements over the last few decades. FTS instruments have many important advantages including: resolving power up to $10^6$ or more, wave number accuracy to a part in $10^7$ or better, broad spectral coverage, and the ability to record all spectral resolution elements in parallel. Unfortunately FTSs also have the disadvantage of multiplex noise compared to a grating spectrometer with an array detector (e.g. Thorne et al. 1999, section 17.1.3). The process of recording and transforming an interferogram leads to a smooth spectral redistribution of the quantum statistical (Poisson) noise from each feature in a spectrum over the entire spectrum. This means that the S/N of a weak line is determined by the photon noise contributed by the strong lines in the spectrum.  In contrast, the S/N of a weak line in a dispersive spectrometer is determined solely by its photon noise.  A cross dispersed and aberration compensated 3m focal length echelle spectrometer was developed in our laboratory astrophysics effort to complement FTS data and provide better results on weak branches, especially in the UV (Wood & Lawler 2012).

In the current study we are focused on UV lines of Co II.  New BF measurements on 84 lines from 20 upper levels are reported.  These new measurements are primarily from cross-dispersed echelle data.  All of the upper levels, except the $z^5D_0$, have LIF radiative lifetime measurements and transition probabilities for stronger lines in this set have been measured and calculated in earlier studies.   This study includes new measurements on many HRLs.

## 2. EMISSION SPECTRA FROM ECHELLE and FTS SPECTROMETERS

Table 1a lists the echelle spectra used in this study.  Spectra 31 – 70 are the same data used in the Co I study (Lawler et al. 2015).  Spectra 71 – 108 are additional new data recorded for this work on Co II.  These latter echelle spectra enable us to measure lines at wavelengths down to 2000 Å.  Three CCD frames are required to cover the width of a complete echelle order in the UV.  Each of the three frames contains lines ranging in wavelength from 2000 Å to 4000 Å.  Lines appearing on two adjacent frames are then used to fix a "bridge" or calibration factors needed to connect BFs from different CCD frames.  Typically a set of five CCD frames is used to provide redundancy and to test for lamp drift.

The last 3 echelle spectra in the list have different settings of the high resolution grating angle to put different pairs of lines on the same CCD frame.  These data were taken to check some unexpected discordance with earlier work by Salih et al. (1985). Shifting the high resolution grating angle to put desired line pairs on a single CCD frame eliminates the need for some bridge factors and any uncertainty of such factors for selected line pairs.

The relative radiometric calibration of the echelle spectrometer is from $D_2$ lamp spectra recorded immediately after each hollow cathode discharge (HCD) CCD frame. The radiometric calibration of the $D_2$ lamp is NIST-traceable. The heavily used $D_2$ lamp is periodically checked against a second identical NIST-traceable $D_2$ lamp with far fewer hours of operation, and against a windowless Ar miniarc (Bridges & Ott 1977) calibrated personally by Dr. Bridges at NIST. The window transmittance of the demountable water-cooled HCD lamp was directly measured. Most of the data on weak HRLs are from 50 mA to 55 mA spectra recorded using the demountable HCD lamp. A reflection technique was used to provide a rough measurement of the window transmittance of the sealed commercial HCD lamps. Although the window of the demountable HCD is Suprasil 2, the transmittance does have some roll off near 2000 Å.[6]

A set of FTS data including three spectra downloaded from the National Solar Observatory (NSO) Digital Library[7] and new FTS spectra recorded at the National Institute of Standards and Technology (NIST) is part of this Co II study as listed in Table 1b. The NSO spectra were recorded using the 1m FTS developed by Brault (1976). The NIST 0.2 m VUV FTS was used to record the new spectra. Although the primary goal of the run on the VUV FTS was to improve hyperfine structure (HFS) data, the two calibrated spectra from the VUV FTS were averaged into the final BF data. Four FTS spectra were used to improve HFS constants for lines of Co II. Measurements on the weak HRLs are primarily from data recorded using the 3m echelle spectrometer as the weak HRLs were typically not detectable in FTS data.

---

[6] Any mention of commercial products within this publication is for information only; it does not imply recommendation or endorsement by NIST.

[7] FTS data are publicly available at http://diglib.nso.edu/.

The calibration of NSO FTS data is described by Lawler (2015). The calibration of the NIST FTS data is based on a different $D_2$ lamp, but the customary method of periodically checking the heavily used NIST $D_2$ lamp against a second identical NIST $D_2$ lamp with far fewer hours of operation is used. The use of HCD lamps running at high current (e.g. in the 500 mA range or more) can yield FTS data on weak HRLs if narrowband filters are introduced to suppress multiplex noise. Unfortunately relative radiometric calibrations over wide wavelength ranges can become rather difficult when narrowband filters are used. The combination of FTS and 3m echelle data makes it possible to measure weak HRLs at modest HCD lamp currents while maintaining broad spectral coverage and high resolving power.

Blends between Co I and Co II lines are separable using the different discharge current dependence of neutral verses ion lines (e.g. Lawler 2011). This separation method works well with a series of spectra recorded using different discharge currents and an identical set up. It does not work as well, however, if the line is blended with another Co II line of excitation. For these lines the availability of very accurate energy levels for Co I (Pickering & Thorne 1996) and for Co II (Pickering et al. 1998) enables us to use a center-of-gravity (COG) technique and resolve blends that Salih et al. (1985) listed as problematic. Blending of dominant lines from two upper levels, the $z^5F^o_2$ at 46452.7 cm$^{-1}$ and the $z^5D^o_2$ at 47537.4 cm$^{-1}$, prevented Salih et al. (1985) from reporting any BFs for those levels. The COG of the blended feature is compared to Ritz wave numbers for both contributing transitions. Then a requirement is imposed that a normalized and weighted combination of the Ritz wave numbers match the COG of the blended feature. This COG method yields the fractional contribution of each line to the blended feature.

For good reliability the COG technique requires a high S/N ratio as well as a Ritz wave number separation of ≈ 0.05 cm$^{-1}$ or more and an accurate wave number calibration of the FTS data. The wave number calibration need not be global and can sometimes be based on nearby unblended strong lines from the same upper level.

Our one standard deviation uncertainties of the branching fraction of unblended lines are estimated from the statistical uncertainty in the measurement of line intensity and the uncertainty in the calibration, estimated as 0.001 % of the wave number difference between the line and the dominant branch from its upper level. Blended lines that have been analyzed with the COG method have an additional uncertainty from the Ritz wave number and the measured COG wave number of the blended line. A combined uncertainty of 0.005 cm$^{-1}$ from the Ritz and measured wave numbers results in a 3 % uncertainty in the intensity ratio and is a typical value for our blended lines. An additional source error is due to optical depth effects in our strongest lines. Optical depth refers to the re-absorption of line radiation by the same line absorption in the HCD lamp measurement. In BF measurements optical depth errors typically first affect the strongest branch from an upper level, especially when that branch connects to the ground or low metastable levels. They yield a small fractional decrease in the apparent BF of the strongest branch, and a larger fractional increase of the apparent BF of weaker branches. Errors due to optical depth can be eliminated from our measurements by using spectra taken at different currents.

3. BRANCHING FRACTIONS

Upper levels in this study have LIF lifetime measurements from Salih et al. (1985) and Mullmann et al. (1998). BFs from these upper levels were also reported by these authors. A comparison to the new BF measurements is presented in Table 2. Salih et al. used FTS data with a relative radiometric calibration from the NO gamma bands. The HCD lamps used for the early FTS measurements were run at currents in the 400 mA to 800 mA range. Mullman et al. used a high resolution echelle spectrometer with a CCD detector array. The HCD lamps used in the early echelle measurements were run at currents in the 80 mA to 700 mA range. The cross dispersed echelle data used in this new work have integrations up to 90 minutes to achieve good S/N on weak lines with lower lamp currents in the 10 mA to 55 mA range.

Each upper level is discussed in a paragraph below. This high level of scrutiny of BF measurements is motivated by the relatively small set of lines, < 100 lines, from only 20 upper levels and by the fact that this is the second set of BF measurements from UW-Madison program for many of the Co II lines. The discordances identified in the paragraphs below are less than the lifetime uncertainties in many cases and thus do not significantly affect the final transition probabilities.

The $z^5F^o_5$ upper level at 45197.7 cm$^{-1}$ has three branches that were all measured by Salih et al. (1985). Our results for this level are primarily based on echelle data and a relative radiometric calibration from a $D_2$ lamp and an Ar miniarc standard lamp. Two of the three branches are in good agreement with Salih et al. but the intermediate strength branch at 2428.3 Å is weaker than that reported by Salih et al. This difference could be explained by some optical depth effects on the strongest line in the older data, however this explanation is not consistent with the good agreement on the strong (BF > 0.97) and weakest (BF < 0.01) branches at 2388.9 Å and 2825.2 Å respectively. The disagreement on the strength of the intermediate branch from this level is the largest discordance (when scaled to reported uncertainties) between our new measurements in Table 2 and those by Salih et al. The discordance has not been fully explained but the possibilities of an impurity in the older data or a typographical error deserve consideration. This discordance did motivate us to make some echelle measurements with the strong and intermediate branch on a single echelle frame as mentioned in the preceding section.

Pickering et al. (1998) report observing six branches from the $z^5F^o_4$ upper level at 45378.8 cm$^{-1}$. One of these branches at 2417.7 Å is labeled as a possible blend with a Co I line. Salih et al. (1985) reported BFs for the strongest five of the six branches observed by Pickering et al. Our results for this level are primarily based on echelle data except for the blended branch. The COG position of the feature in the 600 mA FTS data of Table 1b indicates that the Co I contribution is < 7 % of the entire feature at 2417.7 Å, rising to about 20 % in our 20 mA FTS data. A similar result is found by fitting the HFS using constants derived from other transitions. The BF of the 2417.7 Å line in our 20 mA data agrees with that in our 600 mA data within 8 % when the blend is accounted for using the COG method, which is more reliable than the current dependence method for this blended feature. The 8 % difference is likely due to an optical depth effect from the 2378.6 Å line at 600 mA current. A small correction to the FTS result on this branch is included in our results in Table 2. Satisfactory agreement is found with results from Salih et al. for the other four branches at 2378.6 Å, 2449.2 Å, 2810.9 Å, and 3621.2 Å. (Satisfactory agreement in the context of this work refers to final transition probabilities with overlapping error bars.) The very weak line at 2203.0 Å detected by Pickering et al. is not in our FTS or echelle data.

The $z^5F^o_3$ upper level at 45972.0 cm$^{-1}$ has six branches observed by Pickering et al. (1998) and the strongest four of these six were measured by Salih et al. (1985). The branches at 2383.5 Å and 2834.9 Å are both labeled as Co I blends by Pickering et al. The wave number of the strongest branch with high S/N at 2383.5 Å matches the Co II Ritz wave number in both our 20 mA and 600 mA FTS data of Table 1b and fits the HFS using constants derived from other lines from these levels. The wave number of the weaker branch at 2834.9 Å agrees better with the Co II Ritz wave number than that of the Co I, indicating that the Co I line has a negligible contribution to the intensity in our spectra. Neither of the BFs at 2383.5 Å or 2834.9 Å shows detectable current dependence. Our results for this level in Table 2 are primarily based on echelle data and they agree with Salih et al. for branches at 2383.5 Å, 2414.1 Å, 2437.0 Å, and 2834.9 Å. The very weak line at 3578.0 Å detected by Pickering et al. is not in our echelle data. The other very weak (BF < 0.002) branch at 2220.5 Å is a HRL that connects to the low metastable level at 950.3 $cm^{-1}$.

Pickering et al. (1998) report observing five branches from the $z^5D^o_4$ upper level at 46320.8 $cm^{-1}$, identifying a double classification of the line at 2326.4 Å and an unknown blend of the line at 3501.7 Å. Salih et al. (1985) reported BFs and transitions probabilities for all five branches. The line at 2326.4 Å does not appear to be significantly blended based on its COG position in the FTS data and it can be fitted with a single HFS profile. The results reported in Table 2 are primarily from echelle data. The weak (BF < 0.01) branch at 3501.7 Å is not included due to the unknown blend partner. A fit of the HFS using HFS constants derived from other lines for the Co II line and fake parameters for the unknown blending line indicates that the Co II line is responsible for about 40 % of the observed feature. The omission of the 3501.7 Å feature with a small BF has very little effect on final transition probabilities due the radiative lifetime uncertainty of ±6 %. The four BFs for lines at 2326.5 Å, 2363.8 Å, 2393.9 Å, and 2807.2 Å are in Table 2. The total branching fraction was adjusted to account for the Co I blend at 3501.7 Å by including a residual correction of 0.01 in the calculation of the sum over all the decays from the upper level. There is satisfactory agreement with Salih et al. The strongest branch, as was the case for the $z^5F^o_4$ level discussed above, is slightly suppressed in the older work perhaps from some optical depth.

Pickering et al. (1998) report observing three branches from the $z^5F^o_2$ upper level at 46452.7 cm$^{-1}$, but labeled the strongest branch at 2386.4 Å as a doubly classified line. Salih et al. (1985) did not report BFs and transitions probabilities due to blending of the strongest branch. The blending line is from the $z^1D^o_2$ level at 67209.3 cm$^{-1}$ decaying to the $c^3P_1$ level at 25317.5 cm$^{-1}$. Analysis of the HFS does not help to resolve this blend as we do not have constants derived from other lines for the $z^5F^o_2$ level or either level of the blend. The position of the doubly classified feature in the FTS data indicates that the $c^3P_1$ - $z^1D^o_2$ contribution is < 20 % of the feature in NSO FTS data. The lack of any current dependence in the blended profile compounds the difficulty of using a HFS analysis, even if the constants were available. Although the spin selection rule suggests that the blend should be weak, the calculations of Pickering et al (1998) indicate that only 47 % of the $z^1D^o_2$ is due to the $3d^7(^2P)4p\ ^1D^o_2$ level and that 13 % is due to the $3d^7\ (^2P)4p\ ^3D^o_2$ level. The separation of Ritz wave numbers and the S/N of the combined feature of the two blended lines are large enough to use the COG separation technique on this blend in the NIST VUV FTS data. The results reported here are primarily from echelle data for the two weaker branches at 2408.8 Å and 2423.6 Å and from the FTS data using a COG separation analysis for the strongest branch at 2386.4 Å. The BF uncertainties of the weaker branches are increased by some additional uncertainty of the COG blend separation estimated to be as much as 6 % of the line of interest. The weak (BF < 0.08) branch at 2423.6 Å is a HRL that connects to the metastable level at 5204.7 cm$^{-1}$.

The $z^5F^o_1$ upper level at 46786.4 cm$^{-1}$ has three branches observed by Pickering et al. (1998) and the strongest two of these three were measured by Salih et al. (1985). The branches at 2212.2 Å and 2389.5 Å are labeled as Co II and Co I blends respectively by Pickering et al. The weakest branch at 2212.2 Å is potentially a HRL but blending with the Co II line from the $3d^7(^2H)4d\ ^3I_6$ level at 113803.5 cm$^{-1}$ decaying to the $3d^7(^2H)4p\ ^3I^o_6$ level at 68614.2 cm$^{-1}$ is a concern in BF measurements. Unfortunately the line is not present in our FTS data and thus it is not possible to perform a COG analysis. The 8 eV difference in the lower levels of the blended lines would make the feature usable in stellar spectroscopy. Echelle data provides an upper limit of BF < 0.016 for the branch but it is not included in Table 2 with the two other lines at 2389.5 Å and 2404.2 Å from FTS data only with a COG separation. The COG separation yielded a contribution of ≈ 10 % to the 2389.5 Å blended feature from the Co I line for the 600 mA FTS data and ≈ 20 % for the lower current VUV FTS data from NIST. A residual correction of 0.016 is included for the likely Co II blend at 2212.2 Å. There is good agreement with Salih et al. for this level.

The $z^5D^o_3$ upper level at 47039.1 cm$^{-1}$ has six branches observed by Pickering et al. (1998) and five were measured by Salih et al. (1985). The branches at 2324.3 Å and 4566.6 Å are labeled as a Co II blend and a Co I doubly classified feature respectively by Pickering et al. The position of the strong branch at 2324.3 Å is well separated from other Co II lines at 2324.3208 Å and 2324.2382 Å in our FTS data. The 4566.6 Å feature, identified as $b^1G_4$ - $z^5D^o_3$ in Pickering et al., is almost certainly only a Co I line. The new results in Table 2 for the three strong lines in the 2300 Å range and the two weak branches in the 3430 Å range are primarily from echelle data. All five new measurements are in satisfactory agreement with results from Salih et al.

The $z^5G^o_6$ upper level at 47078.5 cm$^{-1}$ has a single branch 2286.2 Å and our result in Table 2 agrees with Salih et al.

The $z^5G^o_5$ upper level at 47345.8 cm$^{-1}$ has five branches observed by Pickering et al. (1998) and three were measured by Salih et al. (1985). The branches at 2272.2 Å and 5029.5 Å are labeled as a Co II blend and doubly classified feature respectively by Pickering et al. The branch at 2272.2 Å is in our echelle data with BF = 0.012, but it could not be included in Table 2 due to blending. The 5029.5 Å feature is solely a Co II line from the $3d^7(^4F)4f$ (3.5)[4.5]$_5$ level at 110926.3 cm$^{-1}$ decaying to the $3d^7(^4F)4d$ e$^5D_4$ at 91049.4 cm$^{-1}$. The new results in Table 2 for the three lines at 2111.4 Å, 2307.9 Å, and 2663.5 Å are primarily from echelle data. They have a residual correction of 0.012 that is included for the blended branches omitted from the table. The discordance with Salih et al. for the branch at 2663.5 Å is consistent with some optical depth in the older experiment. The very weak (BF < 0.006) branch at 2111.4 Å is a HRL that connects to the ground level.

Pickering et al. (1998) report observing seven branches from the $z^5D^o_2$ upper level at 47537.4 cm$^{-1}$. The weakest branch at 4499.2 Å observed by Pickering et al. is not in our FTS or echelle data. Two of the branches at 2326.1 Å and 2714.4 Å are listed as Co II blends and one branch at 3388.1 Å is described as masked by a much stronger Co I line. Salih et al. (1985) did not report BFs or transition probabilities for lines from this upper level. Four of the seven branches are reported in Table 2 and new results are primarily from echelle data. The blended branch at 2326.1 Å is from FTS data with a COG separation. The COG separation is unusually complicated because the blended feature 2326.1 Å could involve three lines: the $z^5D^o_2$ to a$^5F_3$ decay of interest here, the Co II line from the y$^3G^o_5$ level at 64601.8 cm$^{-1}$ decaying to the a$^3G_5$ level at 21624.5 cm$^{-1}$, and the Co II line from the $3d^7(^4H)5d$ $^5H_7$ level at 111288.0 cm$^{-1}$ decaying to $z^3I^o_6$ level at 68311.1 cm$^{-1}$. Neglecting the LS forbidden line from the high lying 5d $^5H_7$, which has a log($gf$) of only -7 in Kurucz (2011, updated April 8, 2018), yields consistent results for a variety of discharge conditions and is thus given in Table 2. The possibly blended weak line at 2714.4 Å is classified in this work as from the $z^5G^o_3$ level as discussed below based on the COG of the feature in FTS data and based on the level assignments. The feature at 3388.1 Å in our FTS data appears to be almost entirely from a Co I line. However, HFS fitting suggest that there could be an ≈ 1 % contribution from the $z^5D^o_2$ upper level of interest here. If the 3388.1 Å feature has a 1 % contribution from the $z^5D^o_2$ upper level then it would have a BF ≈ 0.01. A residual correction of 0.01 is included in the BF normalization.

Pickering et al. (1998) report observing six unblended branches from the $z^5G^o_4$ upper level at 47807.5 cm$^{-1}$. All six branches are in Table 2 with new measurements primarily from echelle data. Three very weak (BF < 0.003) branches at 2091.1 Å, 2133.5 Å, and 2248.7 Å are all HRLs that connect to the ground 0.0 cm$^{-1}$ and low metastable levels at 950.3 cm$^{-1}$ and 3350.5 cm$^{-1}$. Results for the other three lines at 2283.5 Å, 2311.6 Å, and 2694.7 Å are in good agreement with Salih et al. (1985).

The $z^5D^o_1$ upper level at 47848.8 cm$^{-1}$ has six branches observed by Pickering et al. (1998) and four were measured by Salih et al. (1985). The weakest branch at 4353.7 Å did not appear in our data. The branches at 2918.4 Å and 3352.8 Å are labeled as a doubly classified Co II feature and a unknown blend respectively by Pickering et al. Based on its position in our FTS data, the 2918.4 Å feature is significantly to the blue of the Ritz wave number for the transition of interest, it is also somewhat to the blue of the Ritz value for the line from the $3d^7(^4F)5s\ e^3F_2$ level at 86940.1 cm$^{-1}$ decaying to the $z^3D^o_1$ level at 52684.6 cm$^{-1}$. The 2918.4 Å feature if it was all from the $z^5D^o_1$ upper level of interest would be weak (BF ≤ 0.042) and this branch is neglected in the results of Table 2.  The weak (BF ≈ 0.015) line at 3352.8 Å does not appear blended but is suppressed in Table 2 due to its poor S/N.  The new results in Table 2 for the three strong lines are primarily from echelle data and they have a residual correction of 0.016 included for the weak branches at longer wavelengths.  The strongest two of these lines at 2330.4 Å, and 2344.3 Å, are in satisfactory agreement with Salih et al.

Pickering et al. (1998) observed two unblended branches from the $z^5D^o_0$ level at 47995.6 cm$^{-1}$.  This J = 0 level of the $z^5D^o$ term does not have a LIF radiative lifetime measurement but Salih et al. (1985) reported lifetime measurements of either 3.3 ns or 3.4 ns for the other four levels of the term.   Dominant UV branches yield the short lifetimes of levels in the $z^5D^o$ term and the estimate of 3.3 ns is used here based on the four other lifetimes and the consistency of radiative lifetimes for short-lived levels in a term throughout the Fe-group.  Both lines have BFs in Table 2.

The $z^5G^o_3$ upper level at 48150.9 cm$^{-1}$ has seven branches observed by Pickering et al. (1998).  The extremely weak transitions at 2117.9 Å and 2670.0 Å are not detectable in our data.  The doubly classified line at 3319.2 Å appears to be all a Co I line based on its position.  Although the strongest branch at 2314.1 Å is listed as a blend with the $3d^7(^2P)4d\ ^3F_4$ level at 110725.407 cm$^{-1}$ decaying to the $x^3D^o_3$ at 52684.634 cm$^{-1}$  by Pickering et al., it is sufficiently resolved in our data to yield an accurate BF.  Pickering et al. label the weak line at 2714.4 Å as doubly classified.  As discussed above it should be classified to this $z^5G^o_3$ upper level based on its position and level assignments.    Salih et al. (1985) reported BF measurements for three lines from this upper level.    New echelle measurements on the lines at 2293.4 Å, 2314.1 Å, and 2714.4 Å are in satisfactory agreement with Salih et al.  The very weak (BF < 0.01) branch at 2265.7 Å is a HRL that connects to the low metastable level at 4029.0 cm$^{-1}$.

Pickering et al. (1998) report observing five unblended branches from the $z^5G^o_2$ upper level at 48388.4 cm$^{-1}$. The weakest branch at 2653.1 Å is not detectable in our data. New results primarily from echelle data for the four strongest lines at 2281.0 Å, 2301.4 Å, 2315.0 Å, and 2697.0 Å are included in Table 2. Three of these lines were measured by Salih et al. and are in agreement with their results. The very weak (BF < 0.01) branch at 2281.0 Å is a HRL that connects to the low metastable level at 4560.8 cm$^{-1}$.

The $z^3G^o_5$ upper level at 48556.0 cm$^{-1}$ has four unblended branches observed by Pickering et al. (1998) and all four were measured by Mullman et al. (1998). New results primarily from echelle data are included in Table 2 for all four lines at 2058.8 Å, 2211.4 Å, 2245.1 Å, and 2580.3 Å. Optical depth in the older experiment is a possible cause of differences between the new results and older results. Variations in HFS can produce level to level variations in optical depth. Although the differences between earlier data from Mullman et al. (1998) and our new data are generally small, the new data is thought to be better due to the use of lower discharge currents. Optical depth causes only a small change in the dominant branch but can often substantially enhance weaker BFs because BFs sum to unity be definition.

Pickering et al. (1998) report observing seven branches from the $z^3G^o_4$ upper level at 49348.3 cm$^{-1}$. New results primarily from echelle data for the five lines at 2025.8 Å, 2065.5 Å, 2232.1 Å, 2528.6 Å, and 2587.2 Å are included in Table 2. The very weak (BF < 0.008) branch at 2173.3 Å, which is identified as doubly classified by Pickering et al. is observed in our echelle data and kept in the BF normalization but omitted from Table 2. Some modest differences from the earlier results are apparent in Table 2. The very weak (BF < 0.01) branch at 2205.9 Å is potentially a HRL that connects to the low metastable level at 4029.0 cm$^{-1}$, but it is not adequately resolved in our echelle data.

The $z^3F^o_4$ upper level at 49697.7 cm$^{-1}$ has seven branches observed by Pickering et al. (1998) including one very weak doubly classified line, possibly blended with a Co I line, in the visible at 4500.6 Å. The position of this visible feature in our FTS data is consistent with it being entirely a Co I line. The six unblended branches were measured by Mullman et al. (1998). New results primarily from echelle data are included in Table 2 for the six unblended lines at 2011.5 Å, 2157.0 Å, 2189.0 Å, 2214.8 Å, 2506.5 Å, and 2564.0 Å. Some modest differences from the earlier results are apparent in Table 2.

Pickering et al. (1998) report observing ten branches from the $z^3F^o_3$ upper level at 50381.7 cm$^{-1}$ including one very weak doubly classified line, possibly blended with a Co I line, in the visible at 5211.7 Å and a very weak line at 1984.8 Å (vacuum) below the current wavelength limit of our cross dispersed echelle spectrometer. The optical wavelength and possibly blended branch at 5211.7 Å is detected on our FTS data. It is discarded because the BF is small (< 0.004) even if it is all from the $z^3F^o_3$ upper level. New BF measurements primarily from echelle data on eight remaining branches are all in Table 2 including the lines at 2022.4 Å, 2049.2 Å, 2181.7 Å, 2200.4 Å, 2464.2 Å, 2519.8 Å, 2559.4 Å, and 2693.1 Å. Satisfactory agreement is found with the earlier work by Mullman et al. (1998) except for the weak (BF < 0.02) branch at 2200.4 Å and the very weak (BF < 0.005) branch at 2693.1 Å. The very weak (BF < 0.01) branches at 2049.2 Å and 2181.7 Å are both HRLs connecting to the low metastable levels at 1597.2 cm$^{-1}$ and 4560.7 cm$^{-1}$ respectively.

The $z^3F^o_2$ upper level at 50914.3 cm$^{-1}$ has eight branches observed by Pickering et al. (1998) including one very weak doubly classified line, possibly blended with a Co I line, in the near UV at 3507.8 Å. This possibly blended weak line has a BF < 0.01 and it is omitted from Table 2. A residual correction of 0.013 is included in the BFs of the other seven lines at 2000.8

Å, 2027.0 Å, 2174.9 Å, 2187.0 Å, 2486.4 Å, 2525.0 Å, and 2546.2 Å. Satisfactory agreement is found with the earlier work of Mullman et al. (1998) except for the weakest branch in common. The weakest (BF ≈ 0.014) branch in common at 2000.8 Å and the very weak branch at 2174.9 Å are both HRLs connecting to the low metastable levels at 950 cm$^{-1}$ and 4950 cm$^{-1}$ respectively.

## 4. Co II TRANSITION PROBABILITIES FOR 84 LINES AND COMPARISON TO EARLIER MEASUREMENTS AND CALCULATIONS

The new BF measurements discussed in the preceding section and listed in Table 2 are combined with LIF lifetime measurements from Salih et al. (1985) and Mullman et al. (1998) to determine the absolute transition probabilities of Table 3. Quinet et al. (2016) reported independent measurements of some of the lifetimes from Salih et al. (1985) and Mullman et al. (1998). Of the 12 lifetimes re-measured by Quinet et al., their new measurements agree with the earlier work by Salih et al. and Mullman et al. in 9 cases within the combined error bars. On average the new lifetimes by Quinet et al. are somewhat shorter than those of Salih et al. and Mullman et al. The periodic re-measurement of selected benchmark quality lifetimes as an end-to-end test of the apparatus used by Salih et al. and Mullman et al. is an important quality check. Although the apparatus used by Quinet et al. is described as having better electronic bandwidth and a shorter laser pulse, such claims are not completely equivalent to an end-to-end test of the apparatus. The two benchmark quality lifetimes bracketing the Co II lifetimes of interest are the 1.81 ns lifetime of the resonance $2^1P_1$ level of Be I (Tachiev & Froese Fischer, 1999, estimated to be better than 1 %), and a 3.85 ns lifetime the resonance $3^2P_{3/2}$ level of Mg II (NIST critical compilation of Kelleher & Podobedova 2008; uncertainty of ≤ 1 % at 90 % confidence level).

Figure 1a is a comparison of the new transition probabilities to transition probabilities reported by Salih et al. (1985) and Mullman et al. (1998). Figure 1a is effectively a BF comparison because the same lifetime data are used to normalize the transition probabilities of lines in common. The error bars of Figure 1a are the original transition probability uncertainty combined in quadrature with the new BF uncertainty to avoid double weighting the lifetime uncertainty. As is typical in such plots the dominant branches are concentrated near the horizontal $\Delta \log(gf) = 0$ (perfect agreement line). The discordance on some of the weaker branches (e.g. those with $\log(gf) < -1$) is a reminder of the difficulty of measuring weak branches and establishing reliable error bars on such BFs.

Figure 1b is a comparison of the new transition probabilities to transition probabilities reported by López-Urrutia et al. (1994). These authors used the LIF lifetime measurements from Salih et al. (1985) to normalize some of their BF measurements. López-Urrutia et al. wrote "A 1.33 m McPherson monochromator in Czerny-Turner mounting (grating: 2400 lines/mm) with an EM1 9558 QB photomultiplier tube was used." As in Figure 1a, the error bars of Figure 1b are the original transition probability uncertainty combined in quadrature with the new BF uncertainty to avoid double weighting the lifetime uncertainty. Data points in Figure 1b without error bars are from "preliminary" results reported by López-Urrutia et al. The upward trend for weak branches shown in Figure 1b is an indication of some optical depth in the older experiments. It is standard practice to reduce the discharge current or RF power to test for optical depth problems, but the simultaneous loss of S/N on weaker lines sometimes makes it very difficult to avoid optical depth errors. Figure 1c is a comparison of the new transition probabilities to theoretical transition probabilities reported by Raassen et al. (1998). No error bars are provided on the theoretical results. Nevertheless the comparison is quite encouraging.

It is important to notice that the y-axis or Δlog(*gf*) scale is identical on Figures 1a through 1e. The "orthogonal operator" method used by Raassen et al. appears to work very well with ionized cobalt. The success of the orthogonal operator calculation of transition probabilities indicates that it has captured the level mixing rather well. This is consistent with a private communication from Pickering (2017) in which she acknowledged the importance of Raassen's energy level and transition probability calculations for finding new energy levels and assigning configurations and terms to the new levels found by Pickering et al. (1998). The orthogonal operator method might be useful in addressing the hyperfine structure (HFS) problem for Co II discussed below. The approach used on Co II lines by Lawler et al. (2015), which involved using the FTS profiles to estimate a composite HFS profile, is in need of improvement.

Figure 1d is a comparison of the new transition probabilities to semi-empirical values of the online Kurucz (April 8, 2018) database[8]. No uncertainties are provided in the database, and some of the earlier measurements are included in the database. However, anything lacking with respect to precision in the Kurucz database is easily made up by its comprehensive scale.

Lastly Figure 1e is a comparison of the new transition probabilities to theoretical transition probabilities from Table 6 of Quinet et al. (2016). The results from Quinet et al. are better than those of Kurucz in Figure 1d but not quite as good as those of Raassen in Figure 1c.

5. Hyperfine Structure Constants

---

[8] http://kurucz.harvard.edu/atoms/2701/gf2701.lines

Lines of Co II have hyperfine structure (HFS) due to the I = 7/2 nuclear spin of the only stable $^{59}$Co isotope and this HFS splitting affects the analysis of astrophysical data. The HFS splitting complicates line blending and can de-saturate deep absorption lines that are not on the linear part of the curve-of-growth. Many lines of Co II exhibit significant broadening but unfortunately only a few lines have the principal components resolved in FTS data. Our usual technique for analyzing partially-resolved HFS patterns uses high-accuracy HFS constants measured using Doppler-free laser spectroscopy for a subset of levels as a starting point (Townley-Smith et al. 2016). Once good HFS constants for one level of a transition are known, accurate values for the other level can be derived even if the HFS pattern is only present as an asymmetric broadening of the profile. However, this technique fails in Co II, because no high-accuracy constants derived with laser spectroscopy have been published. Bergemann et al. (2010) mentioned that "A large-scale study of Co II HFS is ongoing at Imperial College, and here we report a detailed investigation of A factors for the $3d^7$ ($^4$P)4s a$^5$P$_{3,2,1}$ levels and for three of the four $3d^7$($^4$F)4p z$^5$D$^o_{4,3,1}$ levels, with immediate application in this work." The eventual publication of this large-scale study will be welcome, but in the interim we have derived HFS constants for a subset of levels giving prominent lines in our stellar spectrum.

Four spectra were analyzed for HFS and are described in Table 1b. Two spectra were taken from the NSO digital archives[6]. These spectra used a high current hollow cathode lamp and HFS patterns of lines from low-lying levels may be affected by optical depth. We thus recorded two additional spectra listed at the end of Table 1b of a low-current commercial lamp using the NIST VUV FTS. The spectra were analyzed with our XGREMLIN software package (Nave et al. 2015), which uses the HFS fitting programs of Pulliam (1977) to determine the HFS constants A (magnetic dipole) and B (electric quadrupole) with their uncertainties. In almost all cases the uncertainty of the B constant was similar to or larger than its value. We have thus not reported B constants in Table 4, although they were included in the fitting procedure.

Our starting point for the analysis was the $3d^7(^4F)4p\ z^5D_0$ level at 47995.6 cm$^{-1}$. This was used to derive the HFS constants for the $3d^7(^4P)4s\ a^5P_1$ level using a transition at 3370.928 Å (29656.9 cm$^{-1}$), shown in Figure 2. The $3d^7(^4P)4s\ a^5P_1$ level was then used to derive HFS constants for the $3d^7(^4P)4p\ z^5S_2$ level, which was in turn used to derive constants for the $3d^7(^4P)4s\ a^5P_{2,3}$ levels. In this way, HFS constants were derived for the 28 levels of Co II in Table 4.

For many of the lines in Co II, the HFS is not resolved and the lines show either a small broadening or slight asymmetry. An example is shown in Figure 3 for the two lines used to derive the HFS A constant for $3d^7(^4P)4s\ a^5F_1$. Since the lines show only a small asymmetry, equally good fits can be obtained with two different HFS A constants for each line. However, only a value of -0.008 cm$^{-1}$ fits both lines equally well. We thus use multiple lines where possible to derive the HFS A constants and remove ambiguity for lines with unresolved structure.

Our program assigns uncertainties to the HFS A constant based on the least squares fit of the line. For lines with unresolved structure, this may underestimate the true uncertainty of the A constant. This is because the HFS A constant is primarily determined by the width of the Doppler broadened line and it is not generally possible to estimate what fraction of the observed width is due to HFS. Our estimates of the HFS A constant for these lines was obtained by fixing the Doppler width at a value similar to that of nearby Co II lines. The uncertainty was then estimated by changing the Doppler width by up to 0.02 cm$^{-1}$, representing a maximum likely uncertainty in the Doppler width, and seeing how the HFS A constant changed. For lines with partially-resolved structure this had little effect on the uncertainty derived from the least squares fit, but for lines with unresolved structure the estimated uncertainty may increase by up to an order of magnitude.

The only previous measurements of HFS in lines of Co II are by Bergemann et al. (2010), who presented HFS A and B constants for the $a^5P$ and $z^5D$ levels. Our values agree within the joint uncertainties. The even parity metastable levels listed in Table 4 typically have larger HFS A constants than the radiating odd parity levels. This is due to the unpaired 4s electron of the even parity levels. The configuration of the ground $a^3F$ term is $3d^8$ and thus the HFS A constants of levels in the ground term are expected to be small.

6. The Cobalt Abundance in the Metal-Poor Main Sequence Turnoff Star HD 84937.

We use the new Co II transition probabilities to re-derive the Co abundance in HD 84937. We are unable to perform a similar analysis for the solar photosphere. Even though six of the transitions considered in this study lie in the accessible optical spectral region, all of them are very strong and nearly every one of them is severely contaminated by lines of other atomic and molecular species.

HD 84937 has served as the test case for our previous papers (listed in §1) on Fe-group species in low metallicity halo stars. HD 84937 is a warm main sequence turnoff star that has good basic observational data, such as accurate broadband colors, accurate parallax and proper motions ($p = 13.74 \pm 0.78$ mas (milliarcsecond, where 1 mas = 4.85 nrad), $\mu_\alpha = 373.05 \pm 0.91$ mas/yr, $\mu_\delta = -774.38 \pm 0.33$ mas/yr; van Leeuwen 2007). These data, combined with many high-resolution spectroscopic studies (e.g., Spite et al. 2017 and references therein) of HD 84937 all have led to well-determined stellar atmospheric parameters. We have employed $T_{eff}$ = 6300 K, log g = 4.0, [Fe/H] = –2.15[9], and $v_t$ = 1.5 km s$^{-1}$ in our papers, very close to the parameters used by Spite et al. Those parameters are adopted in the present work.

The HD 84937 high-resolution data sets are the same as employed previously: (1) a vacuum-*UV* Hubble Space Telescope Imaging Spectrograph (HST/STIS) spectrum, obtained under proposal #7402 (PI: R. C. Peterson), covering the wavelength range 2280 Å $\leq \lambda \leq$ 3120 Å at resolving power R ≈ 25,000 and signal-to-noise values from about 30 to about 60; and (2) an optical near-UV spectrum from the SRO VLT UVES archive, obtained under programs 073.D-0024 (PI: C. Akerman) and 266.D-5655, with 3100 Å $\leq \lambda \leq$ 3900, R ≈ 60,000, and S/N ≈ 100 at 3500 Å.

We consider all 60 transitions with $\lambda \geq$ 2280 Å in Table 3 for possible inclusion in the HD 84937 abundance analysis. Final selection of the lines to be used from this list follows criteria that we have described in detail in previous papers, e.g., Lawler et al. (2013). As before we define a relative line strength factor STR $\equiv \log(gf) - \theta\chi$ where $\chi$ is the excitation energy in eV, and the inverse temperature $\theta$ = 5040/T ≈ 0.8 for HD 84937. Most of the Co II lines at wavelengths $\lambda <$ 2700 Å are strong enough to be detected in the *UV* spectrum of HD 84937. Our search for useful Co II lines with $\lambda \geq$ 2280 Å suggests that those transitions with STR < –2 (48 lines) are strong enough to be detected, while the remaining 12 lines are too weak. Unfortunately the very weak set includes all Table 3 lines in the optical spectral region ($\lambda >$ 3000

---

[9] We use standard abundance notations. For elements X and Y, the relative abundances are written [X/Y] = $\log_{10}(N_X/N_Y)$star – $\log_{10}(N_X/N_Y)_{sun}$. For element X, the "absolute" abundance is written log ε(X) = $\log_{10}(N_X/N_H)$ + 12. Metallicity is defined as [Fe/H].

Å). These lines all have STR < –3, substantially below the detection limit, which we confirm by a failed search to detect them in HD 84937. We estimate that cooler subgiant and giant stars with metallicities [Fe/H] < –2 will have useful Co II lines in their optical near-*UV* spectra.

We compute synthetic spectra for the remaining candidate transitions, and compare them to the HD 84937 spectra described above. The model atmosphere for HD 84937 is interpolated from the Kurucz (2011) grid with the parameters given above. The observed/synthetic spectrum matches are used first to eliminate those lines with severe contamination by other atomic and molecular features, and then to derive Co abundances from the surviving lines. To be consistent with past papers in this Fe-group series and with our earlier series of lab/stellar studies of rare-earth elements (beginning with Lawler et al. 2001) we employ the current version of the LTE line analysis code MOOG[10] (Sneden 1973). We include corrections for scattering in the continuum source functions as implemented by Sobeck et al. (2011). However, since HD 84937 is a relatively warm and high-gravity star, scattering is relatively minor compared to H¯ and H I b-f continuum opacities. Comparisons of synthetic spectra computed with and without continuum scattering show almost no differences even at the shortest wavelengths, $\lambda \approx 2300$ Å. These synthetic spectrum computations result in the elimination of most potential Co II transitions, leaving 19 from which to derive a Co abundance in HD 84937.

Line list assembly for the syntheses is described in detail by Lawler et al. (2013). For 12 Co II lines we have upper and lower level laboratory HFS data (Table 4), as discussed in §5. For another two lines we employ the empirical HFS components developed by Lawler et al. (2015). For the remaining five lines there is no laboratory information with which to compute detailed

---

[10] Available at http://www.as.utexas.edu/~chris/moog.html

HFS patterns or even to estimate HFS broadening in the manner of Lawler et al. However, inspection of our FTS lab spectra reveals that the profiles of all of these lines are very narrow, consistent with single line absorbers or very small HFS splits. These lines can safely be analyzed in our stellar spectrum without considering HFS.

In Table 5 we list the wavelengths, excitation energies in eV, log($gf$) values, and abundances for these lines. The mean abundance is $<\log(\varepsilon)> = 2.67 \pm 0.03$ ($\sigma = 0.11$), and the median value is the same. Adopting a solar Co abundance (Lawler et al. 2015) from Co I lines of $\log(\varepsilon) = 4.96 \pm 0.01$ ($\sigma = 0.06$) yields [Co/H] = –2.28. Then with [Fe/H] = –2.32 ± 0.01 ($\sigma$ = 0.06) from 105 Fe II lines (Sneden et al. 2016) we derive [Co/Fe] = +0.04, essentially the solar abundance ratio. This result is unchanged if we restrict the Sneden et al. Fe II lines to just the 2300 Å to 2600 Å spectral region of our Co II lines; in this case we derive [Fe/H] = –2.32 ± 0.01 ($\sigma = 0.07$) from 45 Fe II lines, yielding the same [Co/Fe] value.

All but the last four lines of Table 5 connect to levels of the two lowest (< 2/3 eV) terms of Co$^+$ and are thus HRLs. Due to the low metallicity of HD 84937, the HRLs of Table 5 are typically not the weak branches from upper levels of interest. Inspection of the Table 5 entries reveals that the non-HRLs (2464.20, 2564.03, 2580.33, and 2587.22 Å) yield relatively small Co abundances in HD 84937. The mean values are $<\log(\varepsilon)>_{HRL} = 2.71 \pm 0.03$ ($\sigma = 0.07$), and $<\log(\varepsilon)>_{nonHRL} = 2.53 \pm 0.05$ ($\sigma = 0.010$). We suspect that the better mean value is that from the HRL set, but to keep consistency with previous papers in this series we will use the mean value from all 19 Co II lines. This issue should be pursued in more detail in the future.

Our new Co/Fe ratio is 0.10 dex smaller than that reported by Lawler et al. (2015) and Sneden et al. (2016). On average the $gf$-values newly reported here are essentially the same as in the earlier study. However, we have added three Co II lines to the Lawler et al. set ($\lambda$ = 2336.23 Å, log($\varepsilon$) = 2.52; 2417.66 Å, 2.74; 2564.03 Å, 2.47) and eliminated one of its lines (2414.07 Å, 2.87) the net effect of which is to decrease the derived Co abundance. Just as importantly, we have solid lab HFS patterns for the majority of our lines, whereas Lawler et al. had only estimated HFS patterns from measurements of the FTS lab data. Finally, we have done all new syntheses with renewed attention to line contamination by non-Co species. The combined effects of all these changes can explain the slightly smaller new Co abundance.

We compare both the old (Sneden et al. 2016) and the newly determined value for Co II in HD 84937 with the recently obtained Cr II value for this star (Lawler et al. 2017). Both of these [Cr/Co] values are illustrated in Figure 4 along with the ratio of these Fe-group elements from the large-scale survey of halo stars from Roederer et al. (2014). While the differences are small between these values, the new abundance ratio for HD 84937 is entirely consistent with the observed ratios in other MP halo stars of similar metallicity. We see the general trend of abundance values versus metallicity (i.e., Galactic chemical evolution) is relatively flat with increasing values and scatter at higher metallicity. These results do not support large overabundances at low metallicities. (Our measurements of Co I and Cr I values show similar results for this star.) We intend to explore further the chemical evolution of the Fe-group elements in the halo stars. Specifically, we will examine whether Cr/Co ratio versus Fe/H shows large variations at the lowest iron abundances, as has sometimes been reported in the past**, or is in fact,** flat over a wide range of low metallicities. We will also explore the positive abundance correlations among the three lightest Fe-group elements Sc, Ti and V to gain insight on advanced core nuclear fusion in massive stars of the early Galaxy.

## 7. Summary

New BF and log(*gf*) measurements are reported for 84 lines of Co II including 12 weak HRLs.  Hyperfine structure A constants for 28 levels of ionized Co from least-square fits to spectral line profiles in FTS data are reported.   The Co abundance in MP halo star HD 84937 is re-determined.  The above results will be useful in future research on Co abundances in stellar photospheres.  The unusually good agreement between the new measurements and computed transition probabilities by Raassen et al. (1998) using the orthogonal operator method is the most interesting result for future lab work.   Current efforts to measure transition probabilities for weak lines of Fe II, including optical and near UV lines accessible to ground based observations, should benefit from the extensive calculations using the orthogonal operator method.

ACKNOWLEDGEMENTS

This work is supported by NASA grant NNX16AE96G (JEL), NASA grant NNH17AE08I (GN),by a Univ. of Wisconsin Hilldale Undergraduate/Faculty Research Fellowship (TF & JEL), and by NSF grant AST1616040 (C.S.).   J.J.C. acknowledges support by the NSF under Grant No. PHY-1430152 (JINA Center for the Evolution of the Elements).

FIGURE CAPTION

Figure 1 panel (a): Comparison of log($gf$) measurements from this work to published measurements by Mullman et al. (1998) open symbols, and to Salih et al. (1985) solid symbols. Error bars from the original work are combined in quadrature with new BF uncertainties. The solid line indicates perfect agreement. Because the same LIF lifetime measurements are used to normalize the new and older BF measurements, this plot can also be interpreted as a BF comparison. Panel (b): same as (a) except a comparison to López-Urrutia et al. (1998). Measurements by López-Urrutia et al. described as preliminary do not have error bars. Panel (c): same as (a) except a comparison to Raassen et al. (1998). Panel (d): same as (a) except a comparison to the online Kurucz (April 8, 2018) database. Panel (e): same as (a) except as a comparison to Quinet et al. (2016).

Figure 2: Hyperfine structure of the $3d^7(^4P)4s\ a^5P_1$ - $3d^7(^4F)4p\ z^5D°_0$ line at 29656.9 cm$^{-1}$. Black solid line and points show the experimental spectrum (third from top in Table 1b). Red line shows a fit of the 3 HFS components.

Figure 3 Left: HFS fits to the transition $3d^7(^4F)4s\ a^5F_1$ - $3d^7(^4F)4s\ z^5D°_2$, for A constants for $a^5F_1$ of 0.035 cm$^{-1}$, 0.0157 cm$^{-1}$ and -0.008 cm$^{-1}$. Right: HFS fits to the transition $a^5F_1$ - $z^5D_1$ for the same A constants for $z^5F_1$. Only a value of -0.008 cm$^{-1}$ fits both lines.

Figure 4: The HD 84937 [Cr/Co] abundance ratio with Cr taken from Lawler et al. (2017) and with Co from Sneden et al. (2016; red filled circle), from this paper (purple filled square), and from the large scale survey by Roederer et al. (2014; light cyan triangles) as a function of metallicity [Fe/H].

Table 1a. Echelle spectra of sealed commercial (≤ 20 mA) and demountable water-cooled (> 20 mA) HCD lamps.

| Index | Date | Serial Numbers[a] | Buffer Gas | Lamp Current (mA) | Wavelength Range (Å) | Resolving Power | Integration Time (minutes) |
|---|---|---|---|---|---|---|---|
| 31 - 35 | 2014 Aug. 8 | 1, 3, 5, 7, 9 | Ar | 10 | 2100 to 3200 | 250 000 | 45, 90, 45, 90, 45 |
| 36 - 40 | 2014 Aug. 19 | 1, 3, 5, 7, 9 | Ar | 10 | 2100 to 3200 | 250 000 | 44, 88, 44, 88, 44 |
| 41 - 45 | 2014 Aug. 20 | 1, 3, 5, 7, 9 | Ne | 15 | 2100 to 3200 | 250 000 | 45, 90, 45, 90, 45 |
| 46 - 50 | 2014 Aug. 22 | 1, 3, 5, 7, 9 | Ne | 20 | 2100 to 3200 | 250 000 | 45, 90, 45, 90, 45 |
| 51 - 55 | 2015 Jan. 17 | 1, 3, 5, 7, 9 | Ar | 10 | 2200 to 3900 | 250 000 | 45, 90, 45, 90, 45 |
| 56 - 60 | 2015 Jan. 19 | 1, 3, 5, 7, 9 | Ar | 15 | 2200 to 3900 | 250 000 | 45, 90, 45, 90, 45 |
| 61 - 65 | 2015 Jan. 15 | 1, 3, 5, 7, 9 | Ne | 15 | 2200 to 3900 | 250 000 | 45, 90, 45, 90, 45 |
| 66 - 70 | 2015 Jan. 16 | 1, 3, 5, 7, 9 | Ne | 20 | 2200 to 3900 | 250 000 | 45, 90, 45, 90, 45 |
| 71 - 75 | 2015 June 4 | 1, 3, 5, 7, 9 | Ar | 10 | 2000 to 2600 | 250 000 | 47, 92, 47, 92, 47 |
| 76 - 80 | 2015 June 5 | 1, 3, 5, 7, 9 | Ar | 15 | 2000 to 2600 | 250 000 | 47, 91, 47, 91, 47 |
| 81 - 85 | 2015 July 3 | 1, 3, 5, 7, 9 | Ne | 20 | 2000 to 2600 | 250 000 | 45, 90, 45, 90, 45 |
| 86 - 90 | 2015 July 9 | 1, 3, 5, 7, 9 | Ne | 20 | 2000 to 2600 | 250 000 | 45, 90, 45, 90, 45 |
| 91 - 95 | 2015 Aug. 29 | 1, 3, 5, 7, 9 | Ar | 50 | 2000 to 2600 | 250 000 | 45, 90, 45, 90, 45 |
| 96- 100 | 2015 Sept. 10 | 1, 3, 5, 7, 9 | Ne | 50 | 2000 to 2600 | 250 000 | 45, 90, 45, 90, 45 |
| 101 -105 | 2015 Oct. 15 | 1, 3, 5, 7, 9 | Ne | 55 | 2000 to 2600 | 250 000 | 45, 90, 45, 90, 45 |
| 106 -108 | 2015 Oct. 28 | 2,5,8 | Ne | 20 | 2000 to 2600 | 250 000 | 90, 90, 90 |

[a]At least three CCD frames are needed to capture a complete echelle grating order in the UV. In the almost all of the above data sets, except the last, five CCD frames are used to provide redundancy and a check for lamp drift. Typically Spectra 1, 5, and 9 have the same echelle grating setting but are analyzed separately. Cumulative integration times of at least 90 minutes were used for all echelle grating settings. The CCD was read-out as needed to avoid saturation. The number of CCD read-outs to achieve 90 minutes varied between 22 and 540 depending on the lamp current and on the strength of spectral lines being observed with an echelle grating setting. Read-time is typically 5 s.

Table 1b. FTS spectra of HCD lamps.

| Use | FTS | Date | Serial Number | Buffer Gas | Lamp Current (mA) | Wavelength Range (Å) | Limit of Resolution ($cm^{-1}$) | Integration Time (minutes) | Beam Splitter | Filter | Detector[a] |
|---|---|---|---|---|---|---|---|---|---|---|---|
| BF | NSO 1m | 1982 Aug. 14 | 5 | Ar-Ne | 600 | 2036 to 13559 | 0.057 | 57 | UV | none | Large UV Si Diode |
| BF | NSO 1m | 1979 Dec. 11 | 3 | Ar | 300 | 2254 to 13378 | 0.058 | 78 | UV | CS9-54 | Super Blue Si Diode |
| BF & HFS | NSO 1m | 1983 Nov. 14 | 6 | Ar-Ne | 600 | 2462 to 6900 | 0.047 | 375 | UV | $CuSO_4$ | Large UV Si Diode |
| BF | NSO 1m | 1982 Aug. 14 | 2 | Ne | 600 | 2269 to 12686 | 0.053 | 58 | UV | none | Large UV Si Diode |
| BF | NIST VUV | 2017 Mar. 21 | 3+4+5 | Ne | 20 | 1786 to 3571 | 0.05 | 176 | $CaF_2$ | none | R7154 PMT |
| BF | NIST VUV | 2017 Mar. 21 | Sum of 7 to 14 | Ne | 20 | 1786 to 3571 | 0.05 | 474 | $CaF_2$ | none | R7154 PMT |

| | | | | | | | | | | |
|---|---|---|---|---|---|---|---|---|---|---|
| HFS | NSO 1m | 1983 Nov. 14 | 3 | Ne | 600 | 2013 to 12566 | 0.057 | 56 | UV | none | Large UV Si Diode |
| HFS | NIST VUV | 2017 Mar. 21 | Sum of 3 to 14 | Ne | 20 | 1786 to 3571 | 0.05 | 741 | CaF$_2$ | none | R7154 PMT |
| HFS | NIST VUV | 2017 Mar. 23 | 2 | Ne | 20 | 2273 to 4545 | 0.05 | 54 | CaF$_2$ | none | R7154 PMT |

Table 2. Branching fractions from this experiment (Expt.) with comparison to measurements by Salih et al. (1985) and Mullman et al. (1998).

| Wavelength in air[a] (Å) | Upper Level[b] (cm$^{-1}$) | J | Lower Level[b] (cm$^{-1}$) | J | Branching Fraction This Expt. | Branching Fraction Other Expt. |
|---|---|---|---|---|---|---|
| 2388.9172 | 45197.708 | 5 | 3350.494 | 5 | 0.977(15) | 0.973(2)[c] |
| 2428.2916 | 45197.708 | 5 | 4028.988 | 4 | 0.0169(10) | 0.026(1)[c] |
| 2825.2369 | 45197.708 | 5 | 9812.859 | 4 | 0.0057(5) | 0.006(1)[c] |
| 2378.6258 | 45378.751 | 4 | 3350.494 | 5 | 0.714(11) | 0.66(1)[c] |
| 2417.6589 | 45378.751 | 4 | 4028.988 | 4 | 0.248(4) | 0.296(6)[c] |
| 2449.1600 | 45378.751 | 4 | 4560.789 | 3 | 0.0244(10) | 0.025(1)[c] |
| 2810.8547 | 45378.751 | 4 | 9812.859 | 4 | 0.0051(4) | 0.0050(7)[c] |
| 3621.2050 | 45378.751 | 4 | 17771.506 | 3 | 0.0088(14) | 0.009(2)[c] |
| 2220.4595 | 45972.033 | 3 | 950.324 | 3 | 0.00181(20) | |
| 2383.4586 | 45972.033 | 3 | 4028.988 | 4 | 0.686(10) | 0.677(5)[c] |
| 2414.0693 | 45972.033 | 3 | 4560.789 | 3 | 0.259(8) | 0.267(5)[c] |
| 2436.9790 | 45972.033 | 3 | 4950.062 | 2 | 0.0491(20) | 0.050(3)[c] |
| 2834.9432 | 45972.033 | 3 | 10708.330 | 3 | .00359(29) | 0.0038(7)[c] |
| 2326.4727 | 46320.829 | 4 | 3350.494 | 5 | 0.239(7) | 0.260(8)[c] |
| 2363.7998 | 46320.829 | 4 | 4028.988 | 4 | 0.722(11) | 0.69(1)[c] |
| 2393.9045 | 46320.829 | 4 | 4560.789 | 3 | 0.0269(8) | 0.033(3)[c] |
| 2807.1759 | 46320.829 | 4 | 10708.330 | 3 | 0.0026(3) | 0.0026(5)[c] |
| | | | | | | |
| 2386.3683 | 46452.697 | 2 | 4560.789 | 3 | 0.635(10) | |
| 2408.7529 | 46452.697 | 2 | 4950.062 | 2 | 0.286(6) | |
| 2423.6240 | 46452.697 | 2 | 5204.698 | 1 | 0.079(3) | |
| | | | | | | |
| 2389.5380 | 46786.406 | 1 | 4950.062 | 2 | 0.526(16) | 0.50(3)[c] |
| 2404.1720 | 46786.406 | 1 | 5204.698 | 1 | 0.458(16) | 0.50(3)[c] |
| | | | | | | |
| 2324.3209 | 47039.102 | 3 | 4028.988 | 4 | 0.273(8) | 0.266(5)[c] |
| 2353.4223 | 47039.102 | 3 | 4560.789 | 3 | 0.666(10) | 0.658(7)[c] |
| 2375.1904 | 47039.102 | 3 | 4950.062 | 2 | 0.0405(8) | 0.044(2)[c] |
| 3415.7682 | 47039.102 | 3 | 17771.506 | 3 | 0.0085(17) | 0.005(3)[c] |
| 3446.3758 | 47039.102 | 3 | 18031.426 | 2 | 0.0127(23) | 0.015(3)[c] |
| | | | | | | |
| 2286.1591 | 47078.491 | 6 | 3350.494 | 5 | 1.000(20) | 1.00[c] |
| | | | | | | |
| 2111.4491 | 47345.842 | 5 | 0.000 | 4 | 0.0055(3) | 0.01(1)[c] |
| 2307.8602 | 47345.842 | 5 | 4028.988 | 4 | 0.874(13) | 0.82(3)[c] |
| 2663.5310 | 47345.842 | 5 | 9812.859 | 4 | 0.108(9) | 0.17(3)[c] |
| | | | | | | |
| 2326.1350 | 47537.362 | 2 | 4560.789 | 3 | 0.306(6) | |
| 2347.3991 | 47537.362 | 2 | 4950.062 | 2 | 0.619(9) | |
| 2361.5201 | 47537.362 | 2 | 5204.698 | 1 | 0.055(4) | |
| 3423.8255 | 47537.362 | 2 | 18338.639 | 1 | 0.0102(20) | |
| | | | | | | |
| 2091.0576 | 47807.490 | 4 | 0.000 | 4 | 0.00191(23) | |
| 2133.4721 | 47807.490 | 4 | 950.324 | 3 | 0.00204(18) | |
| 2248.6676 | 47807.490 | 4 | 3350.494 | 5 | 0.00286(23) | |
| 2283.5215 | 47807.490 | 4 | 4028.988 | 4 | 0.0536(27) | 0.059(6)[c] |
| 2311.6042 | 47807.490 | 4 | 4560.789 | 3 | 0.868(13) | 0.85(2)[c] |
| 2694.6791 | 47807.490 | 4 | 10708.330 | 3 | 0.071(6) | 0.09(2)[c] |
| | | | | | | |
| 2330.3571 | 47848.778 | 1 | 4950.062 | 2 | 0.430(9) | 0.45(1)[c] |
| 2344.2733 | 47848.778 | 1 | 5204.698 | 1 | 0.514(8) | 0.52(2)[c] |
| 3387.6933 | 47848.778 | 1 | 18338.639 | 1 | 0.040(6) | 0.021(5)[c] |
| | | | | | | |
| 2336.2295 | 47995.591 | 0 | 5204.698 | 1 | 0.944(14) | |
| 3370.9224 | 47995.591 | 0 | 18338.639 | 1 | 0.056(10) | |
| | | | | | | |
| 2265.7448 | 48150.937 | 3 | 4028.988 | 4 | 0.00548(22) | |
| 2293.3895 | 48150.937 | 3 | 4560.789 | 3 | 0.0830(17) | 0.103(4)[c] |
| 2314.0565 | 48150.937 | 3 | 4950.062 | 2 | 0.889(13) | 0.87(1)[c] |
| 2714.4413 | 48150.937 | 3 | 11321.859 | 2 | 0.0222(16) | 0.024(5)[c] |
| | | | | | | |
| 2280.9605 | 48388.439 | 2 | 4560.789 | 3 | 0.0081(5) | |
| 2301.4032 | 48388.439 | 2 | 4950.062 | 2 | 0.116(2) | 0.122(5)[c] |
| 2314.9748 | 48388.439 | 2 | 5204.698 | 1 | 0.870(13) | 0.87(1)[c] |
| 2697.0477 | 48388.439 | 2 | 11321.859 | 2 | 0.0065(9) | 0.009(5)[c] |
| | | | | | | |
| 2058.8170 | 48556.049 | 5 | 0.000 | 4 | 0.023(3) | 0.035(2)[d] |
| 2211.4283 | 48556.049 | 5 | 3350.494 | 5 | 0.0198(18) | 0.0273(11)[d] |
| 2245.1289 | 48556.049 | 5 | 4028.988 | 4 | 0.181(14) | 0.195(6)[d] |
| 2580.3263 | 48556.049 | 5 | 9812.859 | 4 | 0.776(16) | 0.743(9)[d] |
| | | | | | | |
| 2025.7596 | 49348.301 | 4 | 0.000 | 4 | 0.043(7) | 0.067(7)[d] |
| 2065.5421 | 49348.301 | 4 | 950.324 | 3 | 0.042(4) | 0.048(5)[d] |
| 2232.0716 | 49348.301 | 4 | 4560.789 | 3 | 0.0339(24) | 0.041(3)[d] |
| 2528.6158 | 49348.301 | 4 | 9812.859 | 4 | 0.472(7) | 0.436(7)[d] |
| 2587.2196 | 49348.301 | 4 | 10708.330 | 3 | 0.393(6) | 0.392(8)[d] |
| | | | | | | |
| 2011.5163 | 49697.680 | 4 | 0.000 | 4 | 0.188(17) | 0.177(8)[d] |
| 2156.9503 | 49697.680 | 4 | 3350.494 | 5 | 0.0040(3) | 0.0038(2)[d] |
| 2188.9993 | 49697.680 | 4 | 4028.988 | 4 | 0.0046(4) | 0.0037(2)[d] |
| 2214.7927 | 49697.680 | 4 | 4560.789 | 3 | 0.0346(28) | 0.041(3)[d] |
| 2506.4644 | 49697.680 | 4 | 9812.859 | 4 | 0.384(12) | 0.384(4)[d] |
| 2564.0344 | 49697.680 | 4 | 10708.330 | 3 | 0.385(12) | 0.383(5)[d] |
| | | | | | | |
| 2022.3538 | 50381.721 | 3 | 950.324 | 3 | 0.197(16) | 0.219(18)[d] |
| 2049.1735 | 50381.721 | 3 | 1597.197 | 2 | 0.0060(5) | |
| 2181.7256 | 50381.721 | 3 | 4560.789 | 3 | 0.00357(25) | |
| 2200.4213 | 50381.721 | 3 | 4950.062 | 2 | 0.0164(13) | 0.0204(15)[d] |
| 2464.1994 | 50381.721 | 3 | 9812.859 | 4 | 0.180(4) | 0.172(4)[d] |
| 2519.8229 | 50381.721 | 3 | 10708.330 | 3 | 0.331(7) | 0.314(10)[d] |
| 2559.4054 | 50381.721 | 3 | 11321.859 | 2 | 0.258(5) | 0.262(9)[d] |
| 2693.0911 | 50381.721 | 3 | 13260.687 | 2 | 0.00401(24) | 0.0024(3)[d] |
| | | | | | | |
| 2000.7930 | 50914.322 | 2 | 950.324 | 3 | 0.0115(13) | 0.007(2)[d] |
| 2027.0404 | 50914.322 | 2 | 1597.197 | 2 | 0.229(21) | 0.251(18)[d] |
| 2174.9217 | 50914.322 | 2 | 4950.062 | 2 | 0.00187(19) | |
| 2187.0389 | 50914.322 | 2 | 5204.698 | 1 | 0.0090(7) | 0.0076(5)[d] |
| 2486.4410 | 50914.322 | 2 | 10708.330 | 3 | 0.189(8) | 0.183(5)[d] |
| 2524.9738 | 50914.322 | 2 | 11321.859 | 2 | 0.529(16) | 0.527(14)[d] |
| 2546.1598 | 50914.322 | 2 | 11651.276 | 2 | 0.0179(9) | 0.0145(9)[d] |

[a]Wavelength values computed from energy levels using the standard index of air from Peck & Reeder (1972).

[b]Energy levels are from the online NIST ASD[5], and are originally from FTS measurements by Pickering et al. (1998).

[c]Salih et al. (1985).

[d]Mullman et al. (1998).

Table 3. Experimental atomic transition probabilities for 84 lines of Co II organized by increasing wavelength in air.

| Wavelength in air[a] (Å) | Upper Level[b] (cm$^{-1}$) | J | Lower Level[b] (cm$^{-1}$) | J | Transition Probability (10$^6$ s$^{-1}$) | log(gf) |
|---|---|---|---|---|---|---|
| 2000.7930 | 50914.322 | 2 | 950.324 | 3 | 4.0 ± 0.5 | -1.92 |
| 2011.5163 | 49697.680 | 4 | 0.000 | 4 | 65. ± 8. | -0.45 |
| 2022.3538 | 50381.721 | 3 | 950.324 | 3 | 68. ± 7. | -0.53 |
| 2025.7596 | 49348.301 | 4 | 0.000 | 4 | 13.0 ± 2.4 | -1.14 |
| 2027.0404 | 50914.322 | 2 | 1597.197 | 2 | 79. ± 9. | -0.61 |
| 2049.1735 | 50381.721 | 3 | 1597.197 | 2 | 2.07 ± 0.24 | -2.04 |
| 2058.8170 | 48556.049 | 5 | 0.000 | 4 | 6.4 ± 0.9 | -1.35 |
| 2065.5421 | 49348.301 | 4 | 950.324 | 3 | 12.8 ± 1.4 | -1.13 |
| 2091.0576 | 47807.490 | 4 | 0.000 | 4 | 0.64 ± 0.09 | -2.43 |
| 2111.4491 | 47345.842 | 5 | 0.000 | 4 | 1.71 ± 0.14 | -1.90 |
| 2133.4721 | 47807.490 | 4 | 950.324 | 3 | 0.68 ± 0.08 | -2.38 |
| 2156.9503 | 49697.680 | 4 | 3350.494 | 5 | 1.37 ± 0.14 | -2.07 |
| 2174.9217 | 50914.322 | 2 | 4950.062 | 2 | 0.64 ± 0.08 | -2.64 |
| 2181.7256 | 50381.721 | 3 | 4560.789 | 3 | 1.23 ± 0.12 | -2.21 |
| 2187.0389 | 50914.322 | 2 | 5204.698 | 1 | 3.1 ± 0.3 | -1.95 |
| 2188.9993 | 49697.680 | 4 | 4028.988 | 4 | 1.57 ± 0.16 | -1.99 |
| 2200.4213 | 50381.721 | 3 | 4950.062 | 2 | 5.7 ± 0.6 | -1.54 |
| 2211.4283 | 48556.049 | 5 | 3350.494 | 5 | 5.5 ± 0.6 | -1.35 |
| 2214.7927 | 49697.680 | 4 | 4560.789 | 3 | 11.9 ± 1.3 | -1.10 |
| 2220.4595 | 45972.033 | 3 | 950.324 | 3 | 0.49 ± 0.06 | -2.60 |
| 2232.0716 | 49348.301 | 4 | 4560.789 | 3 | 10.3 ± 1.0 | -1.16 |
| 2245.1289 | 48556.049 | 5 | 4028.988 | 4 | 50. ± 5. | -0.38 |
| 2248.6676 | 47807.490 | 4 | 3350.494 | 5 | 0.95 ± 0.10 | -2.19 |
| 2265.7448 | 48150.937 | 3 | 4028.988 | 4 | 1.77 ± 0.13 | -2.02 |
| 2280.9605 | 48388.439 | 2 | 4560.789 | 3 | 2.54 ± 0.21 | -2.00 |
| 2283.5215 | 47807.490 | 4 | 4028.988 | 4 | 17.9 ± 1.5 | -0.90 |
| 2286.1591 | 47078.491 | 6 | 3350.494 | 5 | 333. ± 23. | 0.53 |
| 2293.3895 | 48150.937 | 3 | 4560.789 | 3 | 26.8 ± 1.8 | -0.83 |
| 2301.4032 | 48388.439 | 2 | 4950.062 | 2 | 36.2 ± 2.3 | -0.84 |
| 2307.8602 | 47345.842 | 5 | 4028.988 | 4 | 273. ± 17. | 0.38 |
| 2311.6042 | 47807.490 | 4 | 4560.789 | 3 | 289. ± 19. | 0.32 |
| 2314.0565 | 48150.937 | 3 | 4950.062 | 2 | 287. ± 19. | 0.21 |
| 2314.9748 | 48388.439 | 2 | 5204.698 | 1 | 272. ± 17. | 0.04 |
| 2324.3209 | 47039.102 | 3 | 4028.988 | 4 | 80. ± 5. | -0.34 |
| 2326.1350 | 47537.362 | 2 | 4560.789 | 3 | 93. ± 6. | -0.42 |
| 2326.4727 | 46320.829 | 4 | 3350.494 | 5 | 72. ± 5. | -0.28 |
| 2330.3571 | 47848.778 | 1 | 4950.062 | 2 | 126. ± 8. | -0.51 |
| 2336.2295 | 47995.591 | 0 | 5204.698 | 1 | 286. ± 18. | -0.63 |
| 2344.2733 | 47848.778 | 1 | 5204.698 | 1 | 151. ± 9. | -0.43 |
| 2347.3991 | 47537.362 | 2 | 4950.062 | 2 | 188. ± 11. | -0.11 |
| 2353.4223 | 47039.102 | 3 | 4560.789 | 3 | 196. ± 12. | 0.06 |
| 2361.5201 | 47537.362 | 2 | 5204.698 | 1 | 16.6 ± 1.6 | -1.16 |
| 2363.7998 | 46320.829 | 4 | 4028.988 | 4 | 219. ± 13. | 0.22 |
| 2375.1904 | 47039.102 | 3 | 4950.062 | 2 | 11.9 ± 0.8 | -1.15 |
| 2378.6258 | 45378.751 | 4 | 3350.494 | 5 | 204. ± 12. | 0.19 |
| 2383.4586 | 45972.033 | 3 | 4028.988 | 4 | 185. ± 10. | 0.04 |
| 2386.3683 | 46452.697 | 2 | 4560.789 | 3 | 212. ± 14. | -0.04 |
| 2388.9172 | 45197.708 | 5 | 3350.494 | 5 | 279. ± 16. | 0.42 |
| 2389.5380 | 46786.406 | 1 | 4950.062 | 2 | 155. ± 9. | -0.40 |
| 2393.9045 | 46320.829 | 4 | 4560.789 | 3 | 8.1 ± 0.5 | -1.20 |
| 2404.1720 | 46786.406 | 1 | 5204.698 | 1 | 135. ± 8. | -0.46 |
| 2408.7529 | 46452.697 | 2 | 4950.062 | 2 | 95. ± 7. | -0.38 |
| 2414.0693 | 45972.033 | 3 | 4560.789 | 3 | 70. ± 4. | -0.37 |
| 2417.6589 | 45378.751 | 4 | 4028.988 | 4 | 71. ± 4. | -0.25 |
| 2423.6240 | 46452.697 | 2 | 5204.698 | 1 | 26.2 ± 2.1 | -0.94 |
| 2428.2916 | 45197.708 | 5 | 4028.988 | 4 | 4.8 ± 0.4 | -1.33 |
| 2436.9790 | 45972.033 | 3 | 4950.062 | 2 | 13.3 ± 0.9 | -1.08 |
| 2449.1600 | 45378.751 | 4 | 4560.789 | 3 | 7.0 ± 0.5 | -1.25 |
| 2464.1994 | 50381.721 | 3 | 9812.859 | 4 | 62. ± 4. | -0.40 |
| 2486.4410 | 50914.322 | 2 | 10708.330 | 3 | 65. ± 5. | -0.52 |
| 2506.4644 | 49697.680 | 4 | 9812.859 | 4 | 133. ± 10. | 0.05 |
| 2519.8229 | 50381.721 | 3 | 10708.330 | 3 | 114. ± 8. | -0.12 |
| 2524.9738 | 50914.322 | 2 | 11321.859 | 2 | 183. ± 14. | -0.06 |
| 2528.6158 | 49348.301 | 4 | 9812.859 | 4 | 143. ± 9. | 0.09 |
| 2546.1598 | 50914.322 | 2 | 11651.276 | 2 | 6.2 ± 0.5 | -1.52 |
| 2559.4054 | 50381.721 | 3 | 11321.859 | 2 | 89. ± 6. | -0.21 |
| 2564.0344 | 49697.680 | 4 | 10708.330 | 3 | 133. ± 10. | 0.07 |
| 2580.3263 | 48556.049 | 5 | 9812.859 | 4 | 216. ± 13. | 0.37 |
| 2587.2196 | 49348.301 | 4 | 10708.330 | 3 | 119. ± 7. | 0.03 |
| 2663.5310 | 47345.842 | 5 | 9812.859 | 4 | 34. ± 3. | -0.40 |
| 2693.0911 | 50381.721 | 3 | 13260.687 | 2 | 1.38 ± 0.12 | -1.98 |
| 2694.6791 | 47807.490 | 4 | 10708.330 | 3 | 23.8 ± 2.7 | -0.63 |
| 2697.0477 | 48388.439 | 2 | 11321.859 | 2 | 2.0 ± 0.3 | -1.95 |
| 2714.4413 | 48150.937 | 3 | 11321.859 | 2 | 7.2 ± 0.7 | -1.26 |
| 2807.1759 | 46320.829 | 4 | 10708.330 | 3 | 0.79 ± 0.11 | -2.07 |
| 2810.8547 | 45378.751 | 4 | 9812.859 | 4 | 1.46 ± 0.14 | -1.81 |
| 2825.2369 | 45197.708 | 5 | 9812.859 | 4 | 1.62 ± 0.16 | -1.67 |
| 2834.9432 | 45972.033 | 3 | 10708.330 | 3 | 0.97 ± 0.09 | -2.09 |
| 3370.9224 | 47995.591 | 0 | 18338.639 | 1 | 17. ± 3. | -1.54 |
| 3387.6933 | 47848.778 | 1 | 18338.639 | 1 | 11.8 ± 1.9 | -1.21 |
| 3415.7682 | 47039.102 | 3 | 17771.506 | 3 | 2.5 ± 0.5 | -1.51 |
| 3423.8255 | 47537.362 | 2 | 18338.639 | 1 | 3.1 ± 0.7 | -1.57 |
| 3446.3758 | 47039.102 | 3 | 18031.426 | 2 | 3.7 ± 0.7 | -1.33 |
| 3621.2050 | 45378.751 | 4 | 17771.506 | 3 | 2.5 ± 0.4 | -1.35 |

[a]Wavelength values computed from energy levels using the standard index of air from Peck & Reeder (1972).

[b]Energy levels are from the online NIST ASD[5], and are originally from FTS measurements by Pickering et al. (1998).

Table 4. Hyperfine Structure (HFS) A constants for Selected Levels in singly ionized Co.

| Configuration | Term | J | Level (cm$^{-1}$) | HFS A-constant (0.001 cm$^{-1}$) | Levels used in Fitting | Spectra used in Fitting | Previous Measurement[a] (0.001 cm$^{-1}$) |
|---|---|---|---|---|---|---|---|
| 3d$^7$($^4$F)4s | a$^5$F | 5 | 3350.494 | 33.8 ± 0.8 | z$^5$D$_4$ | 1, 3, 4 | |
| | | 4 | 4028.988 | 29.9 ± 0.8 | z$^5$D$_3$ | 1, 3, 4 | |
| | | 3 | 4560.789 | 25.0 ± 0.8 | z$^5$D$_4$, z$^5$D$_3$ | 1, 3, 4 | |
| | | 2 | 4950.062 | 17.6 ± 0.8 | z$^5$D$_3$, z$^5$D$_2$, z$^5$D$_1$ | 1, 3, 4 | |
| | | 1 | 5204.698 | -9.2 ± 1.0 | z$^5$D$_2$, z$^5$D$_1$, z$^5$D$_0$ | 1, 3, 4 | |
| 3d$^7$($^4$F)4s | b$^3$F | 4 | 9812.860 | 2.2 ± 0.8 | z$^5$G$_5$, z$^3$G$_5$, | 3, 4 | |
| | | 3 | 10708.331 | 15.2 ± 0.9 | z$^5$G$_4$, z$^3$G$_4$, z$^3$G$_3$ | 1, 3, 4 | |
| | | 2 | 11321.860 | 50.1 ± 0.9 | z$^3$G$_3$, z$^3$F$_3$ | 1, 3, 4 | |
| 3d$^7$($^4$P)4s | a$^5$P | 1 | 18338.640 | 59.1 ± 0.6 | z$^5$D$_0$ | 1, 2 | 60 ± 2 |
| | | 2 | 18031.427 | 49.4 ± 0.8 | z$^5$S$_2$ | 1, 3 | 49.9 ± 1.5 |
| | | 3 | 17771.507 | 40.2 ± 0.8 | z$^5$S$_2$ | 1, 3 | 40 ± 3 |
| $3d^7(^4F)4p$ | $z^5D^o$ | 4 | *46320.832* | 8.8 ± 0.8 | $a^5P_3, a^5F_4$ | 1, 2, 3, 4 | 9 ± 2 |
| | | 3 | *47039.105* | 7.8 ± 0.8 | $a^5P_3, a^5P_2$ | 1, 2 | 8 ± 20 |
| | | 2 | *47537.365* | 9.0 ± 0.8 | $a^5P_1$ | 2 | |
| | | 1 | *47848.781* | 5.5 ± 0.7 | $a^5P_1$ | 1, 2 | < 10 |
| $3d^7(^4F)4p$ | $z^5F^o$ | 5 | *45197.711* | 10.9 ± 0.8 | $a^5F_5, a^5F_4$ | 1, 3, 4 | |
| | | 4 | *45378.754* | 9.3 ± 0.8 | $a^5F_5, a^5F_4, a^5F_3,$ | 1, 3, 4 | |
| | | 3 | *45972.036* | 11.5 ± 0.8 | $a^5F_4, a^5F_3$ | 1, 3, 4 | |
| $3d^7(^4F)4p$ | $z^5G^o$ | 6 | *47078.494* | 9.9 ± 0.8 | $a^5F_5$ | 3, 4 | |
| | | 5 | *47345.845* | 11.8 ± 0.8 | $a^5F_4$ | 3, 4 | |
| | | 4 | *47807.493* | 15.0 ± 1.0 | $a^5F_3$ | 3, 4 | |
| $3d^7(^4F)4p$ | $z^3G^o$ | 5 | *48556.052* | 13.1 ± 0.8 | $a^5F_5, a^5F_4$ | 1, 3 | |
| | | 4 | *49348.304* | 16.7 ± 0.8 | $b^3F_4, a^5F_3$ | 1, 3, 4 | |
| | | 3 | *50036.348* | 25.6 ± 0.9 | $b^3F_4$ | 1, 3 | |
| $3d^7(^4F)4p$ | $z^3F^o$ | 4 | *49697.683* | 13.6 ± 0.9 | $b^3F_4, b^3F_3$ | 1, 3, 4 | |
| | | 3 | *50381.724* | 18.4 ± 0.9 | $b^3F_4, b^3F_3$ | 1, 3, 4 | |

|  |  | 2 | *50914.325* | 31.3 ± 1.0 | $b^3F_3, b^3F_2$ | 1, 3, 4 |
| $3d^7(^4P)4p$ | $z^5S^o$ | 2 | *56010.475* | -8.7 ± 0.7 | $a^5P_1$ | 1, 3 |

[a]Bergemann et al. (2010)

Table 5. Cobalt Abundances from Co II Lines in HD 84937

| Wavelength in air Å | χ eV | log(*gf*) | log(ε) | HFS[a] |
|---|---|---|---|---|
| 2286.159 | 0.415 | 0.53 | 2.67 | lab |
| 2293.390 | 0.565 | -0.83 | 2.67 | no |
| 2311.604 | 0.565 | 0.32 | 2.67 | lab |
| 2326.135 | 0.565 | -0.42 | 2.77 | lab |
| 2330.357 | 0.613 | -0.51 | 2.79 | lab |
| 2336.230 | 0.645 | -0.63 | 2.52 | no |
| 2353.422 | 0.565 | 0.06 | 2.82 | lab |
| 2361.520 | 0.645 | -1.16 | 2.77 | no |
| 2386.368 | 0.565 | -0.04 | 2.69 | est |
| 2389.538 | 0.613 | -0.40 | 2.72 | no |
| 2393.905 | 0.565 | -1.20 | 2.72 | lab |
| 2404.172 | 0.645 | -0.46 | 2.77 | est |
| 2408.753 | 0.613 | -0.38 | 2.69 | no |
| 2417.659 | 0.499 | -0.25 | 2.74 | lab |
| 2436.979 | 0.613 | -1.08 | 2.67 | lab |
| 2464.199 | 1.216 | -0.40 | 2.52 | lab |
| 2564.034 | 1.327 | 0.07 | 2.47 | lab |
| 2580.326 | 1.216 | 0.37 | 2.44 | lab |
| 2587.220 | 1.327 | 0.03 | 2.67 | lab |

[a] "lab" means that HFS was adopted from the lab results of this paper; "est" means that HFS was taken from the semi-empirical computations reported by Lawler et al. (2015); and "no" means that HFS was not taken into account due to lack of laboratory information.

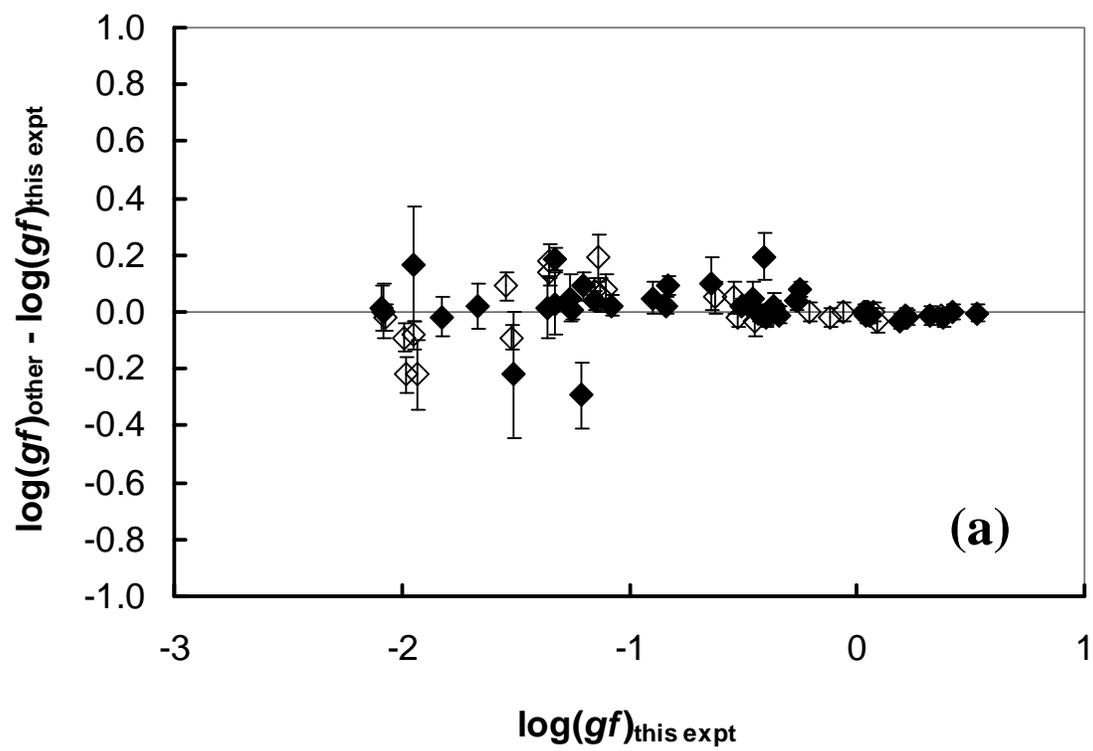
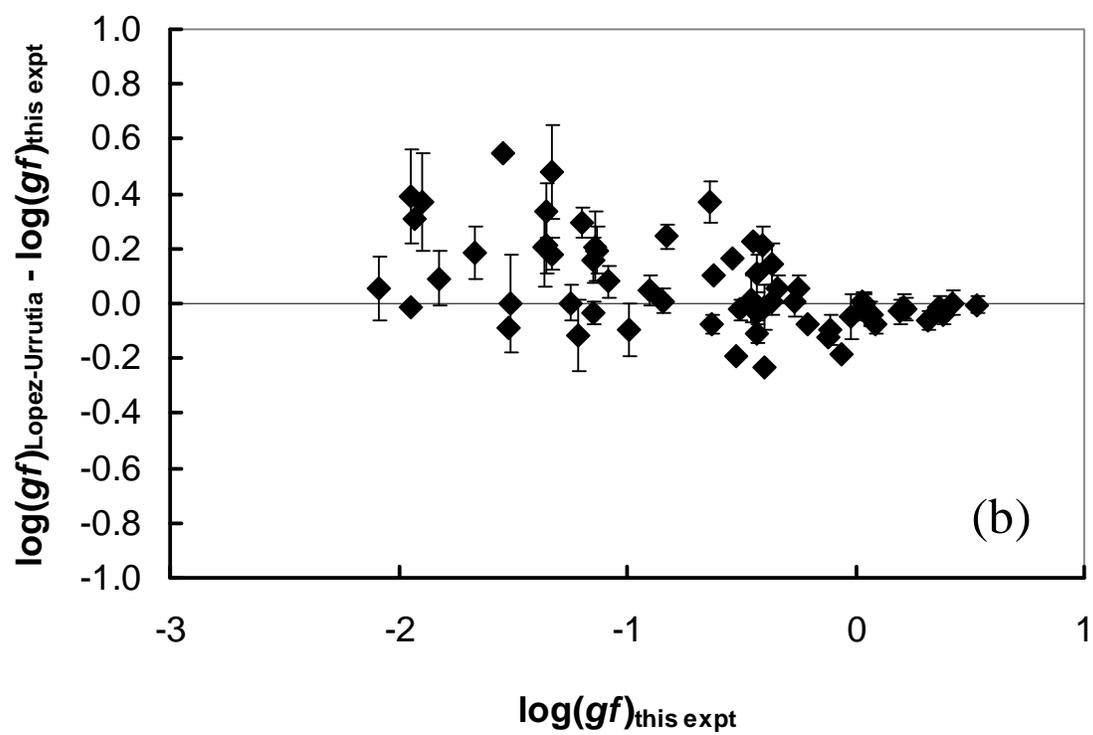

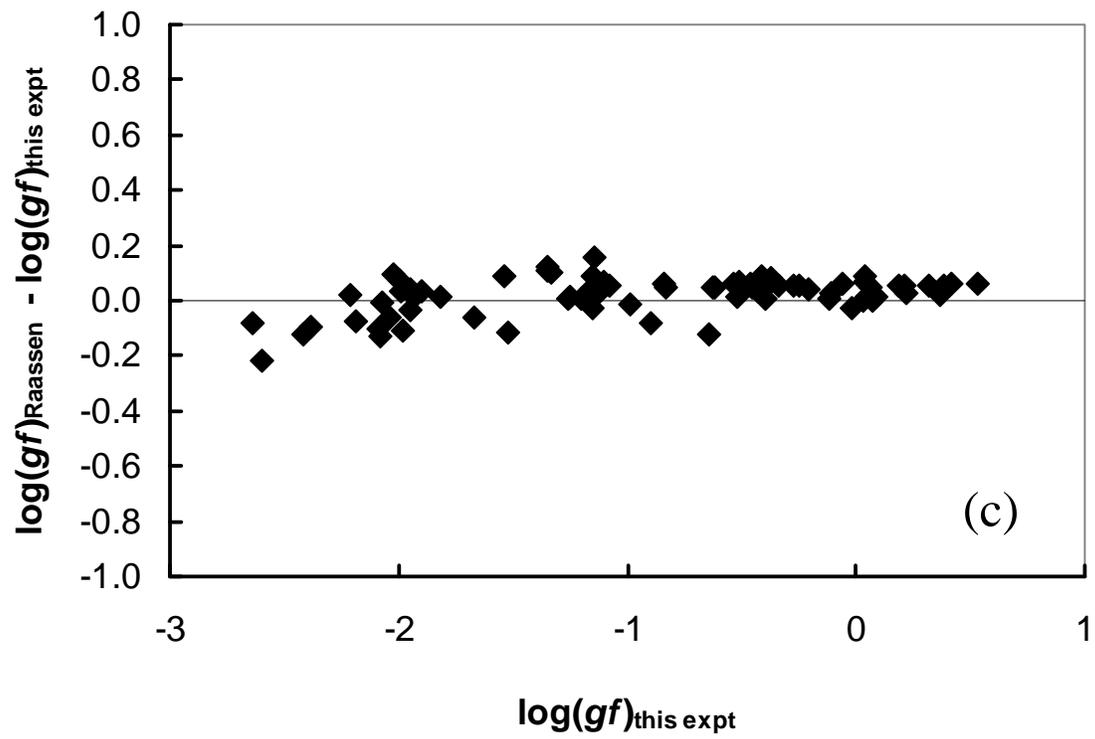
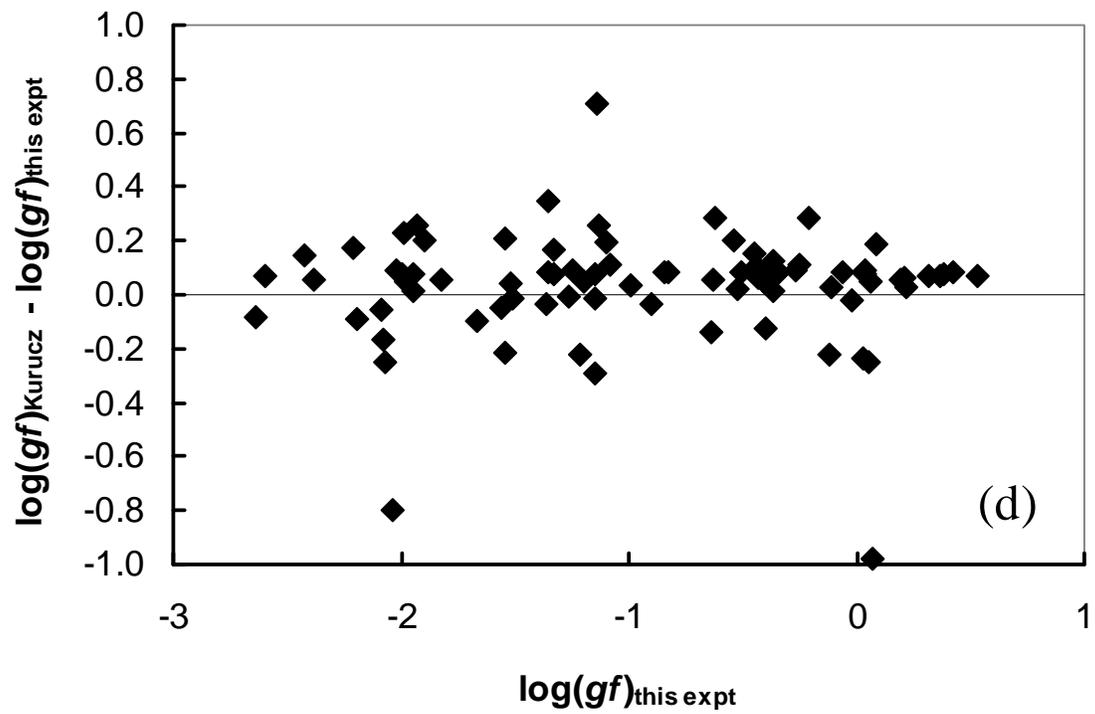

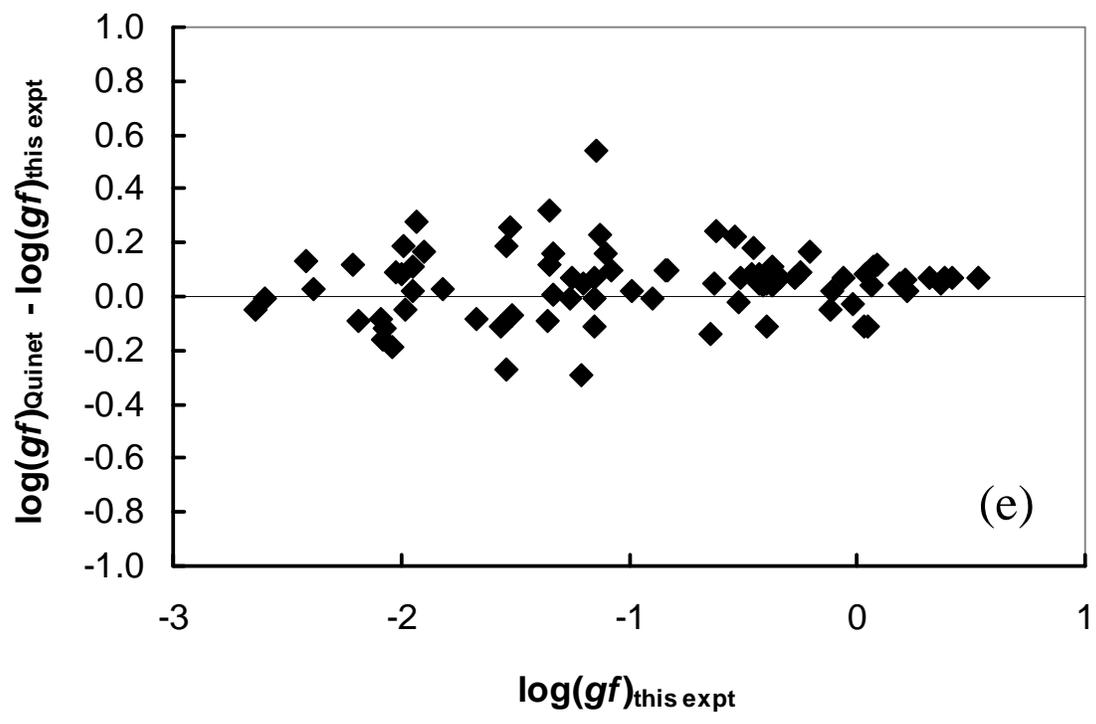

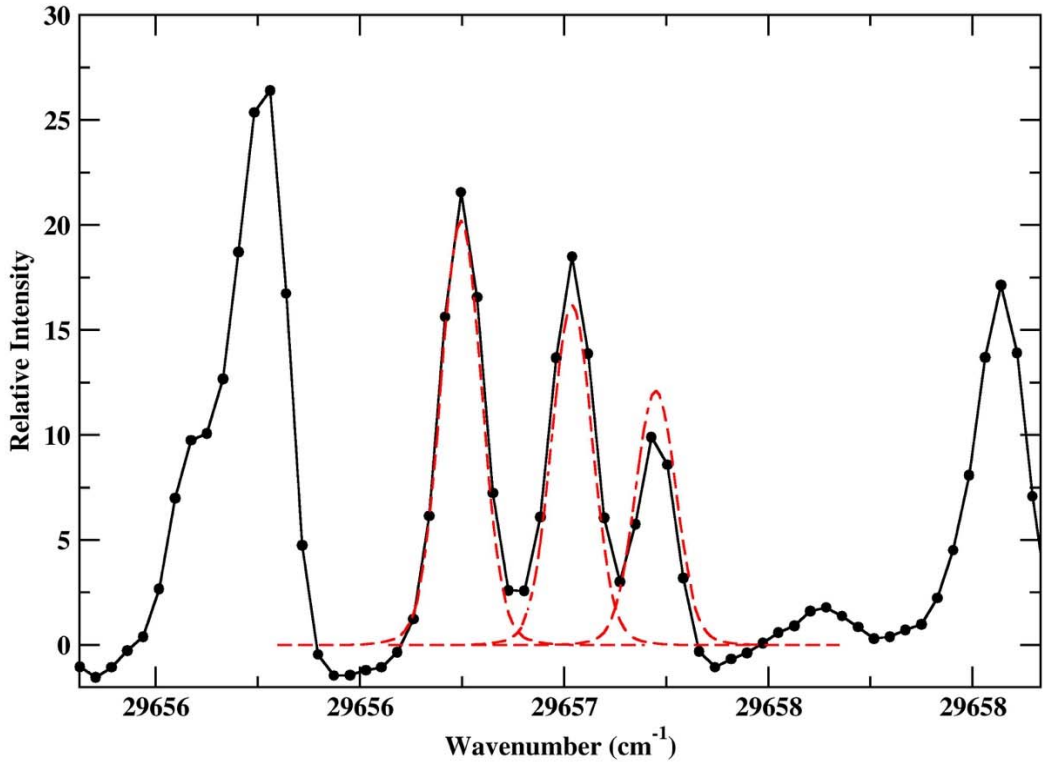

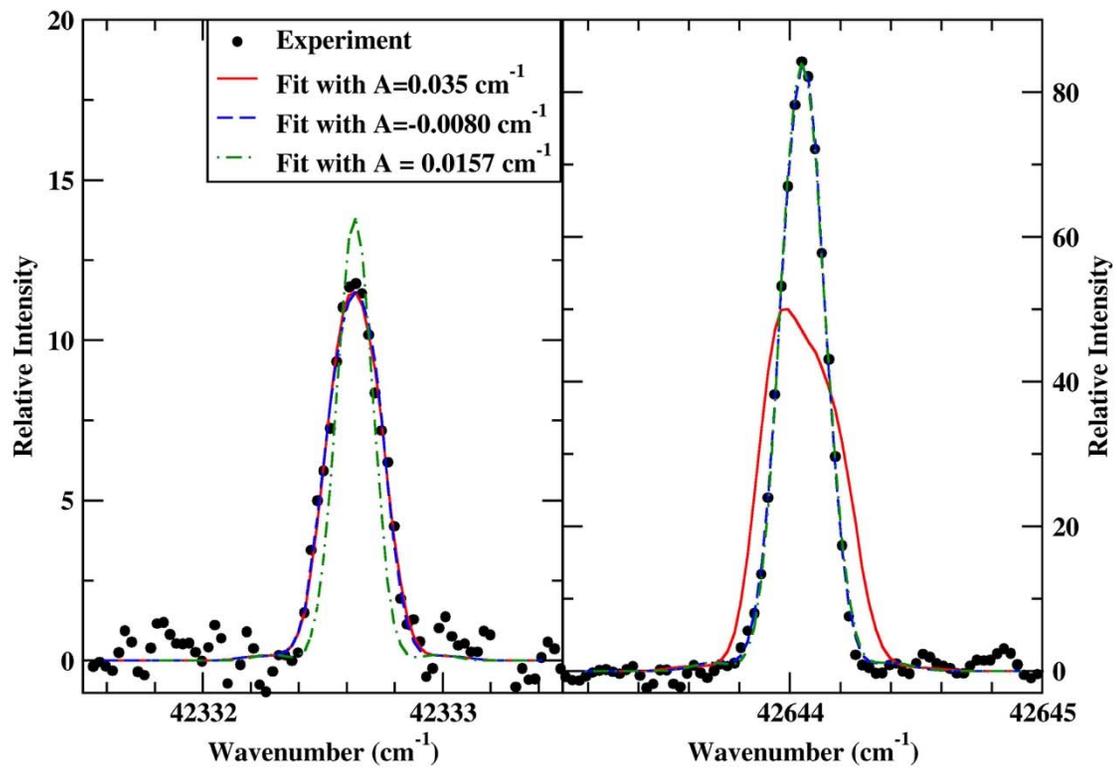

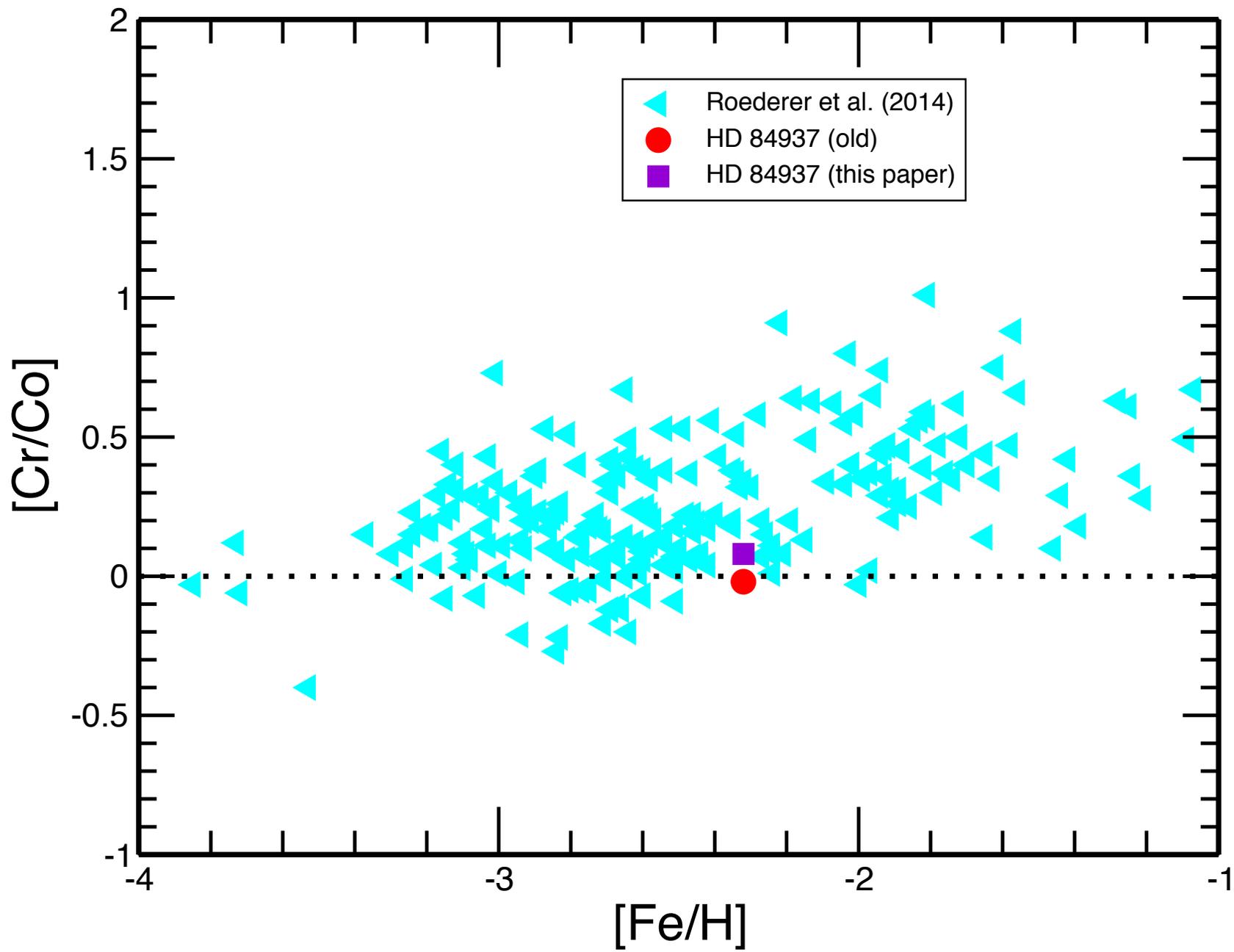